\documentclass[conference,a4paper]{IEEEtran}

\usepackage{enumerate}
\usepackage{amssymb,stmaryrd,amsmath,amsfonts,rotating}%amsfonts,amsthm
\usepackage[noadjust]{cite} \usepackage{color} %temporary packages 
\usepackage{epic}

\usepackage[utf8x]{inputenc}
\usepackage{amsthm}
\usepackage{bbm}

\newcommand{\zz}{\mathbb{Z}}
\newcommand{\bin}{\{\pm 1\}}

\newcommand{\eesym}{\mathbb{E}}
\newcommand{\eepnob}[1]{\eesym_{#1}}
\newcommand{\eenob}{\eesym}
\newcommand{\ee}[1]{\:\eenob\!\left[#1\right]}
\newcommand{\eep}[2]{\:\eepnob{#1}\!\left[#2\right]}
\newcommand{\pch}{p_{Y|X}}
\newcommand{\angbra}[2]{\left\langle #1 \right\rangle_{\!#2}}

\newcommand{\pr}[1]{\:\mathbb{P}\!\left[#1\right]}

\newcommand{\loccode}{\mathcal{C}}

\newcommand{\ldpc}{\mathrm{LDPC}}

\newcommand{\conn}{\mathrm{\bf conn}}
\newcommand{\disc}{\mathrm{\bf disc}}
\newcommand{\coup}{\mathrm{\bf coup}} 
\newcommand{\sigreppar}[1]{\angbra{\sigma^{(1)}_b \cdots\sigma^{(r)}_b}{#1}}
\newcommand{\sigrepgold}{\sigreppar{G}}
\newcommand{\sigrep}{\angbra{\sigma^{(1)}_a \cdots\sigma^{(r)}_a}{}}
\newcommand{\sigrepg}{\angbra{\sigma^{(1)}_a \cdots\sigma^{(r)}_a}{G}}
\newcommand{\dbar}{\bar{d}}
\newcommand{\msf}[1]{\mathsf{#1}}
\newcommand{\xspace}{\mathcal{X}}
\newcommand{\hc}{\mathrm{h}}
\newcommand{\uln}[1]{\underline{#1}}
\renewcommand{\d}[1]{\ensuremath{\operatorname{d}\!{#1}}}
\newcommand{\donly}{\ensuremath{\operatorname{d}}}
\newcommand{\oldx}{U}
\newcommand{\oldy}{V}

\DeclareMathOperator*{\varn}{var}
\DeclareMathOperator*{\pos}{pos}
\DeclareMathOperator*{\occ}{occ}

\newtheorem{lemma}{Lemma}
\newtheorem{theorem}[lemma]{Theorem}
\newtheorem{corollary}[lemma]{Corollary}
\newtheorem{proposition}[lemma]{Proposition}

\begin{document}
\title{Spatial Coupling as a Proof Technique and Three Applications}
%And Now to Something Completely Different: Spatial Coupling as a Proof Technique}

\author{
 \IEEEauthorblockN{Andrei Giurgiu, Nicolas Macris and R\"udiger Urbanke}
 \IEEEauthorblockA{School of Computer and Communication Sciences,\\ EPFL, Lausanne, Switzerland \\
 \{andrei.giurgiu, nicolas.macris, rudiger.urbanke\}@epfl.ch}
}

\IEEEcompsoctitleabstractindextext{
\begin{abstract}

The aim of this paper is to show that spatial coupling
can be viewed not only as a means to build better graphical
models, but also as a tool to better understand uncoupled models.
The starting point is the observation that some asymptotic
properties of graphical models are easier to prove in the case
of spatial coupling. In such cases, one can then use the so-called
interpolation method to transfer known results for the spatially
coupled case to the uncoupled one.

Our main use of this framework is for LDPC codes,
where we use interpolation to show that the average entropy of
the codeword conditioned on the observation is asymptotically
the same for spatially coupled as for uncoupled ensembles. 

We give three applications of this result for a large class of LDPC ensembles. 
The first one is a proof of 
the so-called Maxwell construction stating that the MAP threshold is equal to the Area threshold of the 
BP GEXIT curve. The second
is a proof of the equality between the BP and MAP GEXIT curves above the MAP threshold. The third
application is the intimately related fact that the replica symmetric formula for 
the conditional entropy in the infinite block length limit is exact.

% In a first paper, we have successfully implemented
% this strategy for the case of LDPC ensembles where the variable
% node degree distribution is Poisson. In the current paper we now
% show how to treat the practically more relevant case of general
% variable degree distributions. In particular, regular ensembles
% fall within this framework. As we will see, a number of new technical
% challenges appear, compared to the simpler case of Poisson-distributed
% degrees.

% For our arguments to hold we need symmetry to be present.
% For coding, this symmetry follows from the channel symmetry;
% for general graphical models the required symmetry is called
% Nishimori symmetry.
\end{abstract}
}

\maketitle
\IEEEdisplaynotcompsoctitleabstractindextext
\IEEEpeerreviewmaketitle

\section{Introduction}
Spatially coupled codes were introduced in \cite{felstrom-zigangirov} under the
name of convolutional LDPC codes. It was recently proved
in \cite{shrini-tom-ruedi-universal} that spatial coupling can be used as a paradigm to
build graphical models on which belief-propagation algorithms
perform essentially optimally. The list of applications of this
paradigm has expanded in the past years, to include coding
and compressed sensing, to name two of the most important
ones (see \cite{shrini-tom-ruedi-universal} for a review of history and references). But
spatial coupling can also become useful in a different way: as
a theoretical tool that improves understanding of uncoupled
systems. More specifically, (i) it is sometimes easier to prove
that a property of a graphical model holds under spatial
coupling than it is for the uncoupled version. If that is the case, and
if (ii) the coupled and the uncoupled scenarios are equivalent
with respect to that property, then we obtain a proof that the
uncoupled graphical system has the said property.

In this paper we prove a statement of type (ii) in the case of
LDPC codes. Namely, we prove Theorem \ref{main-thm} which states that the conditional entropy
in the infinite blocklength limit is the same for the coupled
and uncoupled versions of the code. This enables us to derive
the equality of the MAP thresholds for coupled and uncoupled
codes (Corollary \ref{cor-main-thm}). We then present three applications of this result. 
The first one, Equation \eqref{max-construction},
is a proof of the Maxwell construction (see \cite{ruediger-tom-book} Chap 4, Sec. 4.12, p. 257): we already know that this conjecture
holds for coupled ensembles \cite{shrini-tom-ruedi-universal} (a result of type (i)) and here
we deduce that it also holds for the uncoupled systems.
Then, using the freshly-proven Maxwell construction conjecture,
we derive two more results, namely Theorems \ref{th-4} and \ref{th-6}.
The first one states the equality of the BP and MAP GEXIT curves  above the MAP threshold
(see conjecture 1 in \cite{measson} and  Sec III.B \cite{macris07} for a related discussion) and the
second implies the exactness of the replica-symmetric formula for the conditional entropy (see conjecture 1 in \cite{Montanari05tightbounds} and 
Sec III.B in \cite{macris07}). 
Our treatment is general enough to provide a potential recipe for similar results for 
many types of graphical models.

Note that the replica-symmetric formula for error correcting codes on general channels was first derived by non-rigorous methods in the statistical mechanics literature
\cite{17-kabashima-murayama-saad,18-kabashima-murayama-saad-vicente,19-montanari,20-franz-leone-montanari-ricci-tersenghi}. The Maxwell construction and equality of BP and MAP GEXIT curves can also be informally
derived from this formula, which in the statistical physics literature plays the role of a ``more primitive'' object. 
Progress towards a proof of this formula (for general channels) was then achieved 
in the form of a lower bound \cite{Montanari05tightbounds,macris07,macris-kudekar}
and proofs were found that work in low/high noise regimes \cite{21-kudekar-macris} or for the 
special case of the binary erasure channel \cite{22-measson-montanari-urbanke,23-korada-kudekar-macris}.

% [17] Y. Kabashima, T. Murayama, and D. Saad, “Typical Perfor
% mance of Gallager-Type Error-
% Correcting Codes,” Phys. Rev. Lett.
% 84
% , 1355 (2000)

% [18] T. Murayama, Y. Kabashima, D. Saad, and R. Vicente, “Sta
% tistical Physics of Regular
% Low-Density Parity-Check Error-Correcting Codes,” Phys.
% Rev.
% E 62
% , 1577 (2000)

% [19] A. Montanari. “The Glassy Phase of Gallager Codes,” Eur
% . Phys. J. B
% 23
% , 121 (2001)

% [20] S. Franz, M. Leone, A. Montanari, and F. Ricci-Tersengh
% i, “Dynamic phase transition for
% decoding algorithms”, Phys. Rev. E
% 66
% , 046120 (2002)

% [21] S. Kudekar, N. Macris ``Decay of correlations for sparse error correcting codes'', SIAM J. Discrete Math., 25(2), 956–988, July 2011.

% [22] C. Meason, A. Montanari, R. Urbanke “Maxwell Construction: The
% Hidden Bridge between Iterative and Maximum a Posteriori Decoding”,
% IEEE Trans. Inf. Theory, vol 54, no 12 Dec 2008, pp. 5277-5307

% [23] S. B. Korada, S. Kudekar, N. Macris, ``exact solution for the conditional entropy of Poissonian LDPC codes
%over the Binary Erausre Channel'', ISIT June 2007, pp. 1016-1020

Our proof uses the interpolation method,
which was introduced in statistical physics by Guerra and
Toninelli for the Sherrington-Kirkpatrick spin glasses \cite{guerra} and
gradually found its way to constraint satisfaction problems
\cite{franzleone, franzleone2, bayati-gamarnik-tetali} and coding theory \cite{Montanari05tightbounds,macris-kudekar}. The version we use here
employs a discrete interpolation between the coupled and
two versions of the uncoupled scenarios. An error-tolerating
version of the superadditivity lemma is also borrowed from
Bayati \emph{et al.} \cite{bayati-gamarnik-tetali} to show that the conditional entropy has a
limit for large blocklengths (the equivalent of \emph{thermodynamic limit} in
physics terminology).

A \emph{proof of concept} was presented at ISIT 2012
\cite{ours-isit2012} for ensembles with Poisson-distributed degrees,
whose range of applicability in coding is limited. This is due
to the occurence of nodes of very small degrees in significant
proportions, which limits the performance. Here, we
remove this technical barrier and allow for a wide choice of
degree distributions, including regular graphs. However, we
keep the restrictions (see \cite{ours-isit2012}) that the check node degrees
have to be even and that the channel must be symmetric.
The core of the proof rests on the interplay of symmetry and
evenness. A summary of the proof of the main Theorem \ref{main-thm} and the 
application to the proof of the Maxwell construction appeared in ISIT 2013 \cite{ours-isit2013}. The other two applications
presented here are new. 

\section{Preliminaries} \label{sec-prelim}
\subsection{Simple ensembles}
We start by describing the simple (i.e. uncoupled) ensemble of codes, which we denote by $\ldpc(N, \Lambda, K)$, where $N$ 
is the number of variable nodes, $\Lambda(x) = \sum_{d \geq 0} \Lambda_d x^d$ is the probability generating function (PGF) of the variable-node
degree distribution, and the integer $K$ is the fixed check-node degree. The distribution $\Lambda$
must be supported on a finite subset of the positive integers. The average with respect to this
distribution will be denoted by $\dbar$. For each of the $N$ variable nodes, the \emph{target degree}
is drawn i.i.d. from $\Lambda$, and each variable node is labeled with that many \emph{sockets}. The
purpose of a socket is to receive at most one edge from a check node, and all edges must be
connected to sockets on the variable-node side. The number of sockets $D$ will thus be a random variable
which concentrates around $N\dbar$.

The check nodes and the connections are placed in the following way: As long as there are at least $K$ free sockets (initially all sockets are free), add one new check node connected to $K$ free sockets chosen uniformly at random, without replacement. The chosen sockets then become occupied. The final number of check nodes that are added is exactly $\lfloor D/K \rfloor$. Note that there could be at most $K-1$ unconnected sockets at the end of this process, so the resulting variable node degrees will not in general match the target degrees. However, we will be interested in the limit $N \rightarrow \infty$, where the distribution of the resulting degrees matches $\Lambda$.

% COUPLED ENSEMBLES
% Explain that N -> infty, L fixed, K fixed, W fixed

\subsection{Coupled ensembles}\label{subsect-coupled}
Intuitively, a coupled ensemble $\ldpc(N,L,W,\Lambda,K)$ consists of a number $L$ of copies of a simple ensemble, with interaction between copies allowed, in the sense that a check node can be connected to nodes in neighboring copies. More precisely,
the variable nodes are distributed into $L$ groups, which lie on a {\it closed circular chain}. The positions are indexed by integers modulo $L$, and we employ the set of representatives $\{1, \ldots, L\}$. Later we will also refer to open-ended chains.

%The group at position $z$ contains $N$ variable
%nodes, represented by the pairs $(z, 1), \ldots, (z, N)$. Let $V_z$ and $V =
%\cup_z V_z$ denote the set of variable nodes at position $z$, and the set of all
%variable nodes, respectively. If $v \in V_z$, we will also say that $z =
%\pos(v)$.

Just as for simple ensembles, each node is assigned a number of sockets drawn i.i.d. from the
distribution $\Lambda$. The check nodes, however, are restricted in the following way: they are only
allowed to connect to sockets whose positions lie inside an interval - called \emph{window} - of
length $W$ somewhere on the chain, i.e. there exists a position $z$ such that all edges are
connected to check nodes at positions $z, z+1, \ldots, z+W-1$. As before, check nodes have degree $K$, and
they are sampled as follows: first choose a window uniformly at random, then for each edge, choose a
position uniformly and i.i.d. inside that window, and then choose uniformly a free socket at that position. In case there are no free sockets in the chosen position, the process stops. Note that it is possible to stop with a lot of empty sockets in the chain: for example in a very unlucky case, the same position might be picked all the time. However, with high probability, only a small number of sockets will be free at the end of the process, and it is easy to see that in the limit where $N \rightarrow \infty$ the rate of the code only depends on $\dbar$ and $K$. The steps in this process will be described in more detail in Section \ref{sec-conf-model}.
We would like to note that this process is slightly different than the one described in \cite{shrini-tom-ruedi-universal}, but asymptotically equivalent in the large $N$ limit. The reason why we chose this particular coupled construction, which is explained in Section V, is because it is more convenient for the combinatorics entering the interpolation method. The density evolution equations between our construction and the slightly different construction of \cite{shrini-tom-ruedi-universal} are the same. The results of \cite{shrini-tom-ruedi-universal} that we use only rest on the density evolution equations.

The ensembles described so far are built in two
stages: first the vertices are allotted a number of empty sockets,
which is determined by sampling from the distribution $\Lambda$,
thereby establishing the \emph{configuration pattern}; in the second
stage, the edges of the graph are connected to free sockets
in the configuration pattern. It will be sometimes helpful
to separate the two stages and start at the place where the
configuration pattern is already given.

This is a good place to observe that the cases where $W = 1$
and $W = L$ yield instances of the single ensemble in the
following ways: for $W = 1$, there are $L$ different, non-interacting
copies of $\ldpc(N,\Lambda,K)$, whereas for $W = L$,
the whole ensemble is equivalent to $\ldpc(NL,\Lambda,K)$, up to
$O(\sqrt{N})$ missing check nodes.

The reader will notice that the ensemble we have just
constructed is circular and thus the coupling chain has no
boundaries. It is a boundary that is responsible for all the useful
properties of LDPC codes like threshold saturation. We simply
find it easier to work with the circular ensemble and we shall
see later that we can add a boundary condition with little cost.

\subsection{Graphical notation}

Traditionally, the Tanner graph is pictured as a bipartite graph, with edges linking the 
variable nodes to the check nodes. Here we will consider an equivalent rendering, namely 
as a hypergraph, where the variable nodes are the only nodes, and check nodes correspond 
to $K$-ary hyperedges, i.e., $K$-tuples of variable nodes.

The check constraints have fixed even degree $K$, and for each check constraint $a$ we denote by $a_1, \ldots, a_K$
the variables involved in the constraint (the ordering is not important, since we are using this notation to describe a single graph). 
Notation that captures more details will be introduced in Section \ref{sec-conf-model} in order to specify exactly the ensemble of codes.
 For the moment, it suffices to describe a code by listing all of its check constraints, which in turn encode which variables they bind.
 Thus, abusing a bit the standard
terminology, we will say that a graph $G$ is just a $K$-tuple of check constraints
of the kind $a = \{a_1, \ldots, a_K\}$. Note that this notation now allows for repetitions of variables inside check constraints. In general we will use the letters $a$,
$b$, $c$, \dots to describe check constraints, $u$, $v$, \dots to describe
variable nodes, and $G$, $\tilde{G}$ $G'$, \dots to describe graphs.

\subsection{Transmission over channel}
We use these codes to transmit over a binary memoryless symmetric channel
$\pch (y| x)$, where the input symbol set is $\{+1, -1\}$. For just one use of
the channel, it is enough to consider the half-log-likelihood-ratios (HLLR) $h(y)$ instead of
the actual outputs $y$, since they form a sufficient statistic. They are defined (bit-wise)
as
\begin{align} \label{def-hllr}
	h(y) = \frac{1}{2} \log {\pch (y| +1) \over \pch (y| -1)},
\end{align}
with the possibility of taking infinite values. From $h(y)$
one can recover the posterior probability that the bit $x$ was sent. The latter is
easily seen to be proportional to $e^{h(y) x}$. 

A word about notation. In communications it is customary to use $x$ for the channel input and $y$
for the channel output. E.g., we wrote $\pch(y|x)$ for the transition probabilities of the channel, or
we will write $H(\uln X | \uln Y)$ for the conditional entropy of the transmitted vector $\uln X$
given the observed vector $\uln Y$. The methods we apply are from the realm of statistical physics.
In this area it is more common to write $\sigma$ for the input. This notation stems from the fact
that we think of $\sigma$ as a ``spin'' which can take values $\pm 1$.  Hopefully
this causes no confusion. For the output, instead of using $y$ directly, it is slightly more
convenient to 
%We now consider sending the whole input vector, which will be denoted usually by
%$\uln\sigma \in \bin^V$, where $V$ is the set of variable nodes. Instead of the outputs, we
use the HLLRs $\uln h \in (\mathbb{R} \cup \{\pm\infty\})^V$, given by $h_v = h(y_v)$,
where $y$ is the output vector.

The posterior probability that the codeword $\uln\sigma$ was sent, given that $\uln h$ was received, is proportional to
$e^{\uln h \cdot \uln\sigma}$, where $\uln h \cdot \uln\sigma$ stands for the dot product $\sum_{v \in V} h_v
\sigma_v$.  The full expression for the posterior probability
%which holds for non-codewords as well, 
is given by 
\begin{align}\label{eq-gibbs}
	\mu(\sigma) = { e^{\uln h \cdot (\uln \sigma - \uln 1)} \prod_{a \in G}
	\left( 1 + \sigma_a \right)/2 \over Z }, 
\end{align}
where $\sigma_a$ is short for the product $\sigma_{a_1} \cdots \sigma_{a_K}$, and $Z$
is a normalizing factor, also called \emph{partition function}, given by
\[
	Z = \sum_{\sigma \in \bin^V}
	e^{\uln h \cdot (\uln \sigma - \uln 1)} \prod_{a \in G} \frac{ 1 + \sigma_a }{2}.
\]

One can easily check that the product $\prod_{a \in G} (1 + \sigma_a)/2$ is
$1$ when $\uln\sigma$ is any codeword, and $0$ otherwise. The scaling provided by shifting $\uln \sigma$ by $1$ downward
helps to keep the weights involved finite in the case $h = +\infty$. We will see shortly that the case $h=-\infty$ will never occur in our calculations, since by symmetry we can assume the codeword sent is the all-$+1$ codeword.

We have
denoted the above probability measure by $\mu$ in order to distinguish it from other
randomized parameters that appear, notably the channel and the randomness in the
graph $G$. Note that $\mu$ depends on both $G$ and the HLLRs $\uln h$, and when this
is not clear we will make it explicit by adding $G$ or $\uln h$ as a subscript:
$\mu_{G,\uln h}$, $Z(G,\uln h)$. We call this measure the \emph{Gibbs measure}, using a term borrowed from Statistical
Physics. Note that this measure is a random quantity, as it depends on the channel and the random code.

The average with respect to the measure $\mu$ will appear quite often in the
rest of the paper, and we use the \emph{Gibbs
brackets} $\angbra{\cdot}{}$ to indicate it. 
In other words,
\[
	\angbra{f(\uln\sigma)}{} = \sum_{\sigma \in \bin^V} f(\sigma) \mu(\sigma).
\]
Regarding notation, the same subscript conventions, as for $\mu$, apply for the bracket.

Because of symmetry, the channel is characterized by the distribution of the HLLR $h$ computed from
the output of the channel by \eqref{def-hllr} assuming the input of the channel is set to $+1$. We
will view this distribution as a measure $\msf c$ on $\overline{\mathbb{R}} = \mathbb{R} \cup
\{+\infty\}$, which due to channel symmetry has the property
\[
    {\msf c(-h)} = {\msf c(h)} e^{-2h},
    \]

i.e., a $\frac{1}{2}$-log-likelihood $h$ is $e^{2h}$ times more likely to occur than its negative.
For this reason we call this property \emph{symmetry of measures}, we denote all symmetric measures
on $\overline{\mathbb{R}}$ by $\xspace$ and we identify $\xspace$ with the set of BMS channels.

There is a partial ordering, called \emph{degradation}, defined on $\xspace$ which expresses the fact
that one channel is better or worse with respect to another one. We say that a channel $\msf c_1$ is
degraded w.r.t. a channel $\msf c_2$ and write $\msf c_1 \succ \msf c_2$ if there exists a third
channel that can transform the output of $\msf c_2$ (the better channel) into the output of $\msf
c_1$ (the worse channel). For properties of symmetric measures and alternative definitions of
degradedness, we refer the reader to Chapter 4 of \cite{ruediger-tom-book}.
We will denote the \emph{perfect channel}, with respect to which any other BMS channel is degraded, by $\Delta_\infty$. Similarly, the \emph{useless channel},
which is degraded with respect to any other channel, will be denoted by $\Delta_0$.

There are three types of randomness that are involved in our construction: (i) the random graph
which is picked from an LDPC ensemble; (ii) the randomness induced by the channel and (iii) the
Gibbs measure. The expectation in the first case is denoted by $\eesym_{G:\mathcal{G}}[\,\cdot\,]$,
      where $\mathcal{G}$ denotes the ensemble. The expectation with respect to the channel is
      written as $\eesym_h[\,\cdot\,] = \int \cdot \, d\msf c(h)$. As seen before, the average with respect
      to the Gibbs measure is denoted by angular brackets. The symbols
      $\eesym_{G:\mathcal{G}}$ and  $\eesym_h$ commute, since the graph and the channel are
      independent. The angular bracket, however, depends on both $h$ and the graph $G$ and thus
      does not commute with the $\eesym$ symbols. In the language of Statistical Physics, the graph and the channel are said to be quenched. 

There is a deep and useful connection between $\log Z(G,\uln h)$ and the conditional entropy $H(\underline{X}|\underline{Y})$ (where $\underline{X}$ is the input vector and $\underline{Y}$ the output vector). In fact, in our case they are equal, because of the downward shift we added to $\uln\sigma$. 
We would like to express our results in terms of the latter, 
which carries more information-theoretic intuition, but we find it more natural to work with the former. 
\begin{lemma} \label{lem-logz-h}
	For a linear binary code of block length $N$ represented by a graph $G$, we have
%\begin{align}
	\[H(\underline{X}|\underline{Y}) = \eepnob{\uln h}{\log Z(G, \uln h)}.\]
%\end{align}
\end{lemma}
\begin{IEEEproof}
We use successively: (a) the definition of entropy, (b) the fact that apriori all codewords are equally likely to be sent and the symmetry of the channel, which ensures that all terms in the sum are identical, (c) the fact that the log-likelihood is a sufficient statistic, so $p(\uln \sigma|\uln y) = p(\uln \sigma|\uln h)$, and the latter is nothing else than the probability measure $\mu_{G,\uln h}$, and the fact that the distribution of the $\frac{1}{2}$-log-likelihood is given by the distribution $\msf c$ and (d) the fact that $\mu(\uln 1) = Z^{-1}$:
 \begin{align*}
  H(\uln X | \uln Y) &\stackrel{(a)}{=}\!\!\!\! \sum_{\uln \sigma \in \loccode(G)} p(\uln \sigma) \!\int \!\!\int d\uln y \prod_v p_{Y | X}(y_v | \sigma_v) \log p_{\uln X | \uln Y}(\uln \sigma|\uln y)\\
  &\stackrel{(b)}{=}\int \!\! \int d\uln y \prod_v p_{Y | X}(y_v |  1) \log p_{\uln X | \uln Y}(\uln 1|\uln y)\\
  &\stackrel{(c)}{=}\int \!\! \int \prod_v d\msf{c}(h_v) \log \mu_{G,\uln h} (\uln 1)\\
  &\stackrel{(d)}{=}\eepnob{\uln h}{\log Z(G, \uln h)},
 \end{align*}
 where $\loccode(G)$ is the set of codewords.
 
\end{IEEEproof}

\section{Outline of the results} \label{cod:sec2}
\subsection{Comparison of entropies for coupled and simple ensembles}

We will set up the machinery of the interpolation method and direct it at proving the following theorem (for the proof, see Section \ref{sec-large-limit}), which states that the entropies of the simple ${\rm LDPC}(N, \Lambda, K)$ and coupled ${\rm LDPC}(N, L, W, \Lambda, K)$ ensembles are asymptotically the same in the large $N$ limit.
\begin{theorem} \label{main-thm}
	Let $L$, $W$, $K$ be integers such that $L \geq W \geq 1$ and $K$ is even and let $\Lambda$ be a degree distribution with finite support. Then for a fixed BMS channel we have 
\begin{align} \label{eq-limit-simp-coup}
	&\lim_{N \to\infty} \frac{1}{N} \eepnob{G:\ldpc(N,\Lambda,K)}{H(\underline{X}|\underline{Y})} = \nonumber\\
	&\quad\quad=
	\lim_{N \to\infty} \frac{1}{LN} \eepnob{G:\ldpc(N,L,W,\Lambda,K)}{H(\underline{X}|\underline{Y})}, 
\end{align}
and in particular the two limits exist.
\end{theorem}

%TODO: say blocklength entropies
%TODO: what is smooth, cite {measson}
%TODO: useless

The conditional entropy $H(X|Y)$ of the input bit given the output is given in terms of the HLLR
distribution $\msf c$ of the channel for any $\msf c \in \xspace$ by the linear functional
\begin{align}\label{def-entropy-func}
    H(\msf c) = \int \log_2(1 + e^{-2h}) d\msf c(h).
\end{align}

Consider a family of channels $\{\msf c_\epsilon\}$ indexed by a parameter $\epsilon \in
[\underline{\epsilon}, \overline{\epsilon}]$. Such a family is called \emph{smooth} if for all
continuously differentiable functions $f : \overline{\mathbb{R}} \to \mathbb{R}$ such that $f(h)
    e^h$ is bounded, its expectation $\int f(h) d\msf c_\epsilon(h)$ exists and is continuously
    differentiable with respect to $\epsilon$ in $[\underline{\epsilon}, \overline{\epsilon}]$.

A family of channels $\{\msf c_\epsilon\}$ is said to be \emph{ordered by degradation} if $\msf
c_\epsilon \prec \msf c_{\epsilon'}$ whenever $\epsilon < \epsilon'$. For a smooth family of
channels ordered by degradation there is a natural parameterization given by the conditional entropy of the channel. We denote
this special parameter by $\hc$. Starting with any parameter $\epsilon$, we re-parameterize the channel family using $\hc = H(\msf c_\epsilon)$.
It is easy to see that the function $\epsilon \mapsto \hc(\epsilon)$ is continuously differentiable, because $\hc(\epsilon)$ can be obtained by
setting $f(h) = \log_2(1 + e^{-2h})$ in the smoothness property above. Also, the map $\epsilon \mapsto \hc(\epsilon)$ is (by degradation) strictly increasing
and so its inverse $\hc \mapsto \epsilon(\hc)$ is also continuously differentiable. This shows that a family of channels is also smooth with respect to the parameter $\hc$. We will henceforth assume without loss of generality that all channel families are parameterized by the conditional entropy.

Given a smooth family of channels ordered by degradation and parameterized by $\hc$ in the whole
interval $[0,1]$, there exists a value $\hc_{\textrm{MAP}}$ (called the \emph{MAP threshold}) such that for channel parameters below this value, the scaled average conditional entropy (quantities of the kind appearing on both sides of \eqref{eq-limit-simp-coup}) converges to zero in the infinite block length limit, while above this value it is positive. 

More formally, for the two kinds of LDPC ensembles, we define the MAP threshold in the following manner:
%used to be liminf
\begin{align*}
	\hc_{\textrm{MAP}} &= \inf \left\{ \hc : \lim_{N \to\infty} \frac{1}{N} \eepnob{G:\ldpc(N,\Lambda,K)}{H(\underline{X}|\underline{Y})} > 0\right\}, \\ 
	\hc_{\textrm{MAP}}^{L,W} &= \inf \left\{ \hc : \lim_{N \to\infty} \frac{1}{NL} \eepnob{G:\substack{\ldpc\\(N,L,W,\Lambda,K)}}{H(\underline{X}|\underline{Y})} > 0\right\}. 
\end{align*}

These definitions usually employ $\liminf$ and are meaningful even when the existence of limits is not guaranteed. However, in our case, the existence of limits is part of the result of Theorem \ref{main-thm}. The theorem further implies that these two thresholds are equal.

\begin{corollary} \label{cor-main-thm}
	With the same assumptions as in Theorem \ref{main-thm}, we have
    $\hc_{\text{MAP}} = \hc_{\text{MAP}}^{L,W}.$
\end{corollary}

\subsection{Proof of the Maxwell construction}

%A lower bound to this is provided by the so-called \emph{BP threshold}, which is the limit up to which decoding using Belief Propagation is possible. 
As our first application of the equality of MAP thresholds for the coupled and uncoupled ensembles, we will prove the Maxwell conjecture for a large class of degree distributions in the uncoupled case. 

Let us recall the statement of the conjecture. The BP-GEXIT function characterizes asymptotically in
the large $N$ limit an ensemble of codes over a smooth and degraded family of channels and thus is a
function of the channel parameter $\hc$ (see \eqref{eq-def-gbp} for a definition). Supposing now that $\hc$ varies from $0$ to $1$, we define 
 the area threshold $\hc_{\textrm{Area}}$ as that value where
the integral of the BP-GEXIT curve over the interval [$\hc_{\textrm{Area}}, 1]$ equals the design rate $1 - \dbar / K$. 
The \emph{Maxwell construction} conjectures that 
\begin{equation}\label{max-construction}
\hc_{\textrm{Area}} = \hc_{\textrm{MAP}}.
\end{equation}
For more details see \cite{ruediger-tom-book} (Chap 4, Sec. 4.12, pp. 257).

The following was recently proved in \cite{shrini-tom-ruedi-universal}.
For a large class of LDPC ensembles, if we consider the corresponding coupled ensemble, then
the BP threshold (and hence, by threshold saturation, the MAP threshold) is very well approximated  by $\hc_{\textrm{Area}}$ (of the uncoupled ensemble) in the following sense:
\begin{align} \label{eq-maxwell}
	\hc_{\textrm{Area}} - O(\frac{1}{W^{1/2}}) \leq h_{\textrm{BP}}^{L, W, \text{open}}\leq\hc_{\textrm{MAP}}^{L,W,\text{open}} \leq \hc_{\textrm{Area}} + O(\frac{W}{L}).  
\end{align}

The threshold $\hc_{\textrm{MAP}}^{L,W,\text{open}}$ is the one of an \emph{open} coupled chain, which is constructed such that the positions on the chain are from $\{1, \ldots, L\}$, but the windows do not ``wrap around''. 
Instead we add \emph{ghost} variable nodes at positions $-W+2, \ldots, -1, 0$ and $L + 1, \ldots, L
+ W - 1$, whose input bits will always be fixed to $+1$. The windows are of the form $\{z, \ldots, z
+ W - 1\}$, where $z = -W + 2, \ldots, L$.

The only difference in the average conditional entropy of the open and closed chains comes from the
check nodes that lie at the boundary of the chain. The proportion of these check-nodes is $O(W/L)$.
We will later prove in Lemma \ref{lem-z-expl} that the contribution of a single check constraint to the conditional entropy is $O(1)$, and 
so by a repeated application, the difference of the entropies obtained by removing all check constraints on the boundary is $O(W/L)$, which goes to $0$ as $L \to \infty$. As a consequence,
\[\lim_{L \to\infty} \hc_{\textrm{MAP}}^{L,W,\text{open}} = \lim_{L \to\infty} \hc_{\textrm{MAP}}^{L,W}.\]

Thus by \eqref{eq-maxwell} and Corollary \ref{cor-main-thm}, we deduce that in fact $\hc_{\textrm{MAP}}$ equals $\hc_{\textrm{Area}}$, by 
first taking the limit $L \to\infty$ and then $W \to\infty$. This completes the proof that the Maxwell 
construction is indeed correct for all those LDPC ensembles for which \eqref{eq-maxwell} is known.

\subsection{Proof of the equality of the MAP- and the BP-GEXIT curves above the MAP threshold}
%TODO: why partial derivatives
%TODO: german-style prefix splitting
Using the equality of the MAP and area thresholds for uncoupled ensembles, we can derive more properties of uncoupled codes. The ensemble over which we
average in the rest of this section will be exclusively $\ldpc(N,\Lambda,K)$. We first prove the following lemma establishing continuity in the channel parameter for the average per-bit conditional entropy as $N\to\infty$. Also, in order to make clear that the channel output depends
on the channel entropy parameter $\hc$, we will write the former as $Y(\hc)$. 

\begin{lemma}\label{lem-cont-h}
 Given an ensemble $\ldpc(N,\Lambda,K)$ as in Theorem \ref{main-thm} and a smooth family of BMS channels ordered by degradation and parameterized by $\hc$, the quantity $\lim_{N\to\infty} \frac{1}{N} \eep{G}{H(\underline{X} | \underline{Y}(\hc))}$ is a convex function of $\hc$ and is Lipschitz continuous with Lipschitz constant $1$.
\end{lemma}
\begin{IEEEproof}
 That the limit exists and the function is well defined is a consequence of Theorem \ref{main-thm}.
 We use the fact that for any binary linear code the function $\frac{1}{N} H(\underline{X} |
         \underline{Y}(\hc))$ is differentiable and its derivative is increasing with values between $0$ and $1$ \cite[Theorem 5.2, Corollary 5.1]{measson-thesis}, so it is convex and Lipschitz continuous with Lipschitz constant $1$. Taking the average over the code ensemble preserves these two properties. Passing to the limit $N\to\infty$,  Lipschitz continuity and convexity are also preserved, because they are both defined by non-strict inequalities, which are maintained under the pointwise limit.
\end{IEEEproof}

The MAP-GEXIT function
$g^{\textrm{MAP}}$ is defined \cite[Definitions 3 and 6]{measson} as 
%\begin{align}\label{eq-def-gmap}
%	g^{\textrm{MAP}} (\hc) = \limsup_{N\to\infty} \frac{1}{N} \eep{G}{\sum_v H(X_v | Y_{\sim v}(\hc))},
%\end{align}
%where $\sim\! v$ represents the set of all nodes \emph{except} $v$.
%Equivalently \cite[Section III]{measson}, the MAP-GEXIT function also takes the form
\begin{align}\label{eq-equiv-gmap}
    g^{\textrm{MAP}} (\hc) = \limsup_{N\to\infty} \frac{1}{N} \eep{G}{\frac{\donly}{\d \hc} H(\underline{X} | \underline{Y}(\hc))}.
\end{align}

%The derivative ${\frac{d}{d \hc} H(\underline{X} | \underline{Y}(\hc))}$ for any binary linear code along a smooth BMS channel family parameterized by $\hc$ is strictly increasing and takes values in the $[0,1]$ interval \cite[Theorem 5.2, Corollary 5.1]{measson-thesis}.
%Thus the conditional entropy per bit $\frac{1}{N}H(\underline{X} | \underline{Y}(\hc))$ is a convex function with slope bounded by 1. Performing the average over the code ensemble, these properties are preserved. Passing to the limit when $N\to\infty$ 

%and so is the limit of these functions as $N\to\infty$. This also establishes the continuity of $\lim_{N\to\infty} \frac{1}{N} \eep{G}{H(\underline{X} | \underline{Y}(\hc))}$ as a function of $\hc$.

We lower bound the area below $g^{\textrm{MAP}}$ above the MAP threshold as follows:
\begin{align} \label{eq-map-r}
    &\int_{\hc_{\textrm{MAP}}}^1 g^{\textrm{MAP}}(\hc) \d\hc = \nonumber\\
	&\quad{=} \int_{\hc_{\textrm{MAP}}}^1   \left( \limsup_{N\to\infty} \frac{1}{N}
            \eep{G}{\frac{\donly}{\d \hc} H(\underline{X} | \underline{Y}(\hc))} \right)\d\hc\nonumber\\
    &\quad\stackrel{(a)}{\geq} \limsup_{N\to\infty} \int_{\hc_{\textrm{MAP}}}^1  \frac{1}{N} \eep{G}{\frac{\donly}{\d \hc} H(\underline{X} | \underline{Y}(\hc))} \d\hc \nonumber\\
    &\quad\stackrel{(b)}{=} \lim_{N\to\infty} \left(\frac{1}{N} \eesym_G H(\underline{X}|\underline{Y}(1)) - \frac{1}{N} \eesym_G H(\underline{X}|\underline{Y}({\hc_{\textrm{MAP}}}))\right) \nonumber\\
	&\quad\stackrel{(c)}{=} R - 0 = R,
\end{align}
where in step (a) we use the Fatou Lemma (note that the integrand on the r.h.s. is bounded), in step (b) we integrate and then use the existence of limits provided by Theorem \ref{main-thm} to replace $\limsup$ with $\lim$, and in step (c) we observe the following. For the first term, since at $\hc=1$ the channel is completely useless, we have that $H(\underline{X} | \underline{Y}(1)) = H(\underline{X})$, which when scaled by $N$ is nothing else than the rate of the code; in the large blocklength limit, the average of this over the ensemble coincides with the design rate $R = 1 - \bar{d}/K$. For the second term, note that $\lim_{N\to\infty} \frac{1}{N} \eep{G}{H(\underline{X} | \underline{Y}(\hc))} = 0$ which follows from the of continuity in $\hc$ obtained in Lemma \ref{lem-cont-h}.

The BP-GEXIT curve is defined \cite[Definition 6]{measson} by
% We now define the BP-GEXIT curve in a similar way, except that now in the calculation of the ``extrinsic'' entropy of bit $X_v$ we only consider the information in the computation tree of depth $\ell$ rather than all of $Y_{\sim v}$. Denoting the subset of outputs occurring in the computation tree of length $\ell$ by $Y_{\sim v}$ (see \cite[Section III]{measson}), we can formalize the previous statement as
\begin{align}\label{eq-def-gbp}
	&g^{\textrm{BP}} (\hc) = \lim_{\ell\to\infty} \limsup_{N\to\infty} \frac{1}{N} \eep{G}{\sum_v 
        g^{\textrm{BP}}_{G,v}(\hc)},\\
    &g^{\textrm{BP}}_{G,v} (\hc) = \frac{\partial H(X_v | Y_v(\hc_v),
            \Phi^\ell_v(\hc))}{\partial\hc_v} \Big\vert_{\hc_v = \hc},
\end{align}
where $\Phi^\ell_v(\hc)$ is the BP estimate of $X_v$ based on a computation tree of depth $\ell$. An
equivalent form is given by Equation \eqref{eq-49} in Appendix \ref{app-b}.

It is known that (see Lemma 9 in \cite{measson})
\begin{align} \label{eq-eq-map-bp}
g^{\textrm{MAP}} (\hc) \leq g^{\textrm{BP}} (\hc), \text{ for all $\hc \in [0,1]$.}  
\end{align}

The area threshold mentioned before is defined as the solution $\hc_{\textrm{area}}$ to the equation
\begin{align} \label{eq-bp-r}
\int_{\hc_{\textrm{area}}}^1 g^{\textrm{BP}}(\hc) \d\hc = R. 
\end{align}

Using then the equality of the MAP and area thresholds established in the previous subsection for the above-mentioned class of LDPC codes and using \eqref{eq-map-r} and \eqref{eq-bp-r} we obtain

\begin{align}
\int_{\hc_{\textrm{MAP}}}^1 (g^{\textrm{BP}}(\hc) - g^{\textrm{MAP}}(\hc)) \d\hc \leq R - R = 0.
\end{align}

The positivity of the integrand (cf. \eqref{eq-eq-map-bp}) entails the following result.

\begin{theorem} \label{th-4}
    Given an $\ldpc(N, \Lambda, K)$ ensemble and a smooth family of channels indexed by the entropy
    parameter $\hc$, the two curves $g^{\textrm{MAP}}$ and  $g^{\textrm{BP}}$ are equal almost
    everywhere above the MAP threshold, as long as the MAP threshold is at least $\bar{\hc}$ defined in Lemma \ref{lem-cond-tech} below.\footnote{The value $\bar{\hc}$ will always be under the MAP threshold as long as degree are large enough.}
\end{theorem}

The discussion of \eqref{eq-map-r} also entails the following result, which will be useful
subsequently. Among others, this allows us to exchange the $\liminf$ with $\lim$ in the expression
for the MAP threshold.

\begin{proposition}
    The limit $\lim_{N\to\infty} \frac{1}{N}
\eepnob{\ldpc(N,\Lambda,K)}{H(\underline{X}|\underline{Y}(\hc))}$ exists for all values of $\hc$,
    and furthermore
   \[
    \int_{\hc_0}^1 g^{\textrm{MAP}}(\hc) \d\hc = R - \lim_{N\to\infty}\frac{1}{N}
\eepnob{\ldpc(N,\Lambda,K)}{H(\underline{X}|\underline{Y}(\hc_0))},
   \]
   where $R = 1 - \Lambda'(1)/K$ is the rate of the code.
\end{proposition}

\subsection{Exactness of the replica-symmetric formula}

\newcommand{\opc}{\circledast}
\newcommand{\ops}{\boxast}

The previous result, namely the equality of the BP and MAP GEXIT curves, allows us to settle another
conjecture. We can prove that under certain conditions (above the MAP threshold) the potential
functional \cite{pfister-nicolas-kumar-young}, \cite{kumar2014threshold},
           also called replica-symmetric functional, is in fact equal to the conditional entropy
$H(\underline{X}|\underline{Y})$. Note that while the former is a quantity derived by message
passing, the latter is related to combinatorial optima. Also, unlike GEXIT curves, these quantities
make sense already without considering the channel as part of a smooth family and thus in a sense appear to be
more natural. 
% This section contains a result which is very similar to Lemma 26 from
% \cite{shrini-tom-ruedi-universal}. We offer a somewhat different proof for the sake of completeness.

In order to define the potential functional (or replica-symmetric functional), we need to introduce the density evolution operations.
The beliefs that are transmitted during BP have distributions that are symmetric measures. We use
two operations that act on measures, $\opc,\ops : \xspace \times \xspace \to \xspace$, which
correspond to the operations carried on beliefs in BP: the measure $\msf z_1 \opc \msf z_2$ is the
distribution of the sum of two independent random variables 
$$
h_1 + h_2
$$
with laws $h_1\sim\msf z_1$ and $h_2\sim \msf z_2$,
respectively; the measure $\msf z_1 \ops \msf z_2$ is the distribution of 
$$
\tanh^{-1}(\tanh h_1
        \tanh h_2),
$$ 
where $h_1 \sim \msf z_1$ and $h_2 \sim \msf z_2$ are independent random
variables. The operations can be generalized straightforwardly to apply to any finite signed
measures. 

The two operations on measures are commutative, associative and bilinear with respect to 
addition of measures. By $\msf z^{\opc n}$ we mean 
the $\opc$-product of $\msf z$ with itself $n$ times. Given a 
polynomial $\lambda(u) = \sum_{n = 0}^{\deg\lambda} \lambda_n u^n$, we define 
$\lambda^{\opc}(\msf z) = \sum_{n = 0}^{\deg\lambda} \lambda_n \msf z^{\opc n}$. The definitions 
of $\msf z^{\ops n}$ and $\lambda^{\ops}(\msf z)$ are similar.

We restrict ourselves now to regular LDPC ensembles with left and right degrees $d_l$ and $d_r$,
   respectively. However, since the derivation holds more generally, we will work with the
   polynomials $\Lambda, P$ and $\lambda, \rho$ as left and right degrees from the node and from the
   edge perspective, respectively. For us, they take the simple forms $\lambda(u) = u^{d_l-1}$,
   $\rho(u) = u^{d_r-1}$, $\Lambda(u) = u^{d_l}$ and $P(u) = u^{d_r}$.

The density evolution (DE) equation can then be written as $\msf x^{\ell + 1} = \msf c \opc
\lambda^{\opc}(\rho^{\ops}(\msf x^\ell))$. The fixed point that can be reached by starting with $\msf
x^0 = \Delta_0$ will be called \emph{forward DE fixed point} and will be denoted by $\msf x_{\msf c}$.

We are now ready to define the replica-symmetric functional, which depends on the channel $\msf c$ and the message density $\msf x$ as
\begin{align}\label{eq-15}
 \Phi(\msf x, \msf c) &= - \frac{\Lambda'(1)}{P'(1)} H(P^{\ops}(\msf x)) - \Lambda'(1)H(\rho^\ops(\msf x)) 
 \nonumber\\
         &\!\!\! + \Lambda'(1) H(\msf x \ops \rho^{\ops}(x)) + H(\msf c \opc \Lambda^\opc (\rho^\ops (\msf x))).
\end{align}
For a more complete exposition of this formalism, the identity of the potential functional and the replica symmteric functional properties, and various
properties of the two operations $\opc$ and $\ops$,
please refer to \cite{kumar2014threshold}
(note that $\Phi(\msf x, \msf c)$ is equal to minus the function $U(\msf x, \msf c)$ of 
reference \cite{kumar2014threshold}).

The replica-symmetric formula conjectures that 
\begin{equation}\label{replic-sym-form}
 \lim_{N\to +\infty} \frac{1}{N} \eesym_{\ldpc(N,d_l,d_r)}
    H(\underline{X}|\underline{Y}(\msf c)) = \sup_{\msf x\in \xspace} \Phi(\msf x, \msf c).
\end{equation}
We prove this conjecture for standard regular LDPC codes with large enough, but fixed, $d_l$, $d_r$ and also require 
even $d_r$. The proof of this conjecture is a consequence of Theorem \ref{th-6} below.

This theorem states that in a region of channels above the MAP threshold characterized by a
regularity condition, this functional evaluated at the right fixed point (which is algorithmic in nature as it comes from message
        passing) is equal to the conditional entropy, which is combinatorial in nature. 
%         In other
% words, the Bethe approximation of the entropy is exact in the said region.

To express the regularity constraint, we first define the region of channels above the MAP
threshold:
\[
    \mathcal{C}_0 = \{\msf c \in \xspace : \lim_{N\to\infty} \frac{1}{N} \eesym_{\ldpc(N,d_l,d_r)}
    H(\underline{X}|\underline{Y}(\msf c)) > 0.\}
\]

Ideally, we would like our result to hold in the whole of this region, but, unfortunately, we need
to add a Lipschitz type of restriction. Let

\begin{align}\label{eq-16}
 \mathcal{C}_1 &= \big\{ \msf c_0 \in \xspace : \text{there is $\delta > 0$ s.t. for all } \msf c,
     \msf c' \in [\msf c_0, \msf \Delta_0]  \nonumber\\
         &\quad \text{we have that } \left| \frac{\mathcal{B}(\msf x_{\msf c} - \msf x_{\msf c'})}
     {\mathcal{B}(\msf c - \msf c')} \right| \leq \frac{1}{\delta} \big\},
\end{align}
 where 
     \[
     [\msf c_0, \msf \Delta_0] = \{\msf c: \msf c = p \msf c_0 + (1-p)\msf \Delta_0, \text{ for some $p \in
         [0,1]$}\}
     \]
and $\mathcal{B}(\cdot)$ is the Bhattacharyya functional (see  \cite{ruediger-tom-book} for details) defined for any $\zz \in \xspace$ by
\begin{align}\label{eq-50}
\mathcal{B}(\zz) = \int \zz(h) e^{-h} d\zz(h).
\end{align}

Note that the regions $\mathcal{C}_0$ and $\mathcal{C}_1$ depend on the parameters of the code.
\begin{theorem} \label{th-6}
 Given the regular ensemble $\ldpc(N, d_l, d_r)$ with even $d_r$, for any channel $\msf c \in
     \mathcal{C}_0 \cap \mathcal{C}_1$ we have that
\[
    \Phi(\msf x_{\msf c}, \msf c) = \lim_{N\to\infty} \frac{1}{N} \eesym_{\ldpc(N,d_l, d_r)}
 H(\underline{X}, \underline{Y}(\msf c)).
     \]
 \end{theorem}

 As the proof is fairly technical, we defer it to section \ref{app-b}. 
 
 We show now that for large
     degree pairs, $\mathcal C_0 \subseteq \mathcal C_1$, i.e. the theorem holds everywhere above
     the MAP threshold. This is made precise by Lemma 18 from \cite{shrini-tom-ruedi-universal},
         reproduced below. It states that all channels are in $\mathcal C_1$, at least when their entropy is above a certain value. Moreover, this value tends to
             $0$ as the right-degree increases.

\begin{lemma} \label{lem-cond-tech}
 Let $d_l$ and $d_r$ be fixed numbers. There is a constant\footnote{An expression for $\bar{\hc}$
     can be found in Lemma 18 of \cite{shrini-tom-ruedi-universal}.} $\bar{\hc}$ depending only on
     the degrees $d_l$ and $d_r$ satisfying 
     \begin{align}\label{eq-17}
            \bar{\hc} < \frac{e^{1/4} \sqrt{2}}{d_r^{1/4}}
     \end{align}
     such that $\{\msf c\in \xspace : H(\msf c) > \bar{\hc}\} \subseteq \mathcal C_1$.
 \end{lemma}

We can readily see that for large degrees the right hand side of condition \eqref{eq-17} approaches
    $0$. Also, for large degrees, the MAP threshold approaches capacity and is bounded away from $0$
    uniformly for all channel families. This implies that $\mathcal C_0 \subseteq \mathcal C_1$ and hence 
    $\mathcal C_0 \cap \mathcal C_1 = \mathcal C_0$. 
    
We believe that the theorem remains true without this technical condition. Proving that this is
    indeed the case is an interesting open problem.
    
    Let us conclude this paragraph by remarking that the above considerations imply 
    the replica-symmetric formula \eqref{replic-sym-form} for large enough $d_l, d_r$ and where $d_r$ is an even number.
    From \cite{Montanari05tightbounds} we know that (for any BMS channel and $d_r$ even)
    \begin{equation}
     \lim_{N\to\infty} \frac{1}{N} \eesym_{\ldpc(N,d_l,d_r)}
    H(\underline{X}|\underline{Y}(\msf c))
    \geq \sup_{\msf x\in \xspace} \Phi(\msf x, \msf c).
    \end{equation}
    Note first that for $\msf c \notin \mathcal C_0$ we have by definition 
    $\lim_{N\to\infty} \frac{1}{N} \eesym_{\ldpc(N,d_l,d_r)}
    H(\underline{X}|\underline{Y}(\msf c)) = 0$. Thus 
    \begin{equation}
     0\geq \sup_{\msf x\in \xspace} \Phi(\msf x, \msf c)\geq \Phi(\Delta_\infty, \msf c) = 0,
    \end{equation}
so \eqref{replic-sym-form} is satisfied for $\msf c \notin \mathcal C_0$.
    Now consider $\msf c \in \mathcal C_0$.
    Whenever
 $\mathcal C_0 \cap \mathcal C_1 = \mathcal C_0$ (e.g when $d_l, d_r$ are large enough) Theorem \ref{th-6} implies 
    \begin{align}
    \Phi(\msf x_c, \msf c) & =  \lim_{N\to\infty} \frac{1}{N} \eesym_{\ldpc(N,d_l,d_r)}
    H(\underline{X}|\underline{Y}(\msf c))
    \nonumber \\ &
    \geq \sup_{\msf x\in \xspace} \Phi(\msf x, \msf c) 
    \nonumber \\ &
    \geq \Phi(\msf x_{\msf c}, \msf c)
    \end{align}
and hence again \eqref{replic-sym-form} holds for $\msf c\in \mathcal C_0$. 

\section{Some useful lemmas} \label{cod:sec3}

We present in this section two results that are quite general in nature, meaning that they are true
for any linear code. They already appear in \cite{Montanari05tightbounds, macris07}, but we
reproduce short proofs here in order to make the exposition self-contained. The symmetry of the channel
is a property that seems indispensable for the proofs in the rest of this paper, and we will need it
in the form of the Nishimori Identity. The channel used for transmission needs to be BMS, symmetry being the crucial ingredient.

\begin{lemma}[Nishimori Identity] \label{lem-nishi}
Fix a graph $G$ (no constraints on the check node degrees needed here) and a channel $\msf c \in
\xspace$. For any
odd positive integer $m$ we have
\begin{align} \label{eq-nishi}
	\eep{h}{\angbra{\sigma_b}{}^{m}} = \eep{h}{\angbra{\sigma_b}{}^{m+1}},
\end{align}
where $b = (b_1, \ldots, b_J)$ is a vector of variable nodes (which need not belong a check
constraint) of arbitrary length, and $\sigma_b = \sigma_{b_1}\cdots\sigma_{b_J}$. 
\end{lemma}

\begin{IEEEproof} 
We will assume here that the measure $\msf c$ does not contain mass at infinity. Extending to the
general case  can easily be done by considering the point mass at $+\infty$ separately. Because of
channel symmetry, the measure defined by $d\msf s(\msf h) = e^{-h} d\msf c(h)$ has the property $d\msf s(h)
    = d\msf s(-h)$. Using the memoryless property of the channel, the l.h.s.  of
\eqref{eq-nishi} can be written as

\begin{align} 
	\eep{h}{\angbra{\sigma_b}{}^{m}} = \int \angbra{\sigma_b}{}^{m}\prod_{v \in V} 
     e^{h_v} \d{\msf s}(h_v).
\end{align}

 We now observe that due to channel symmetry the above quantity is preserved
 under the transformation 
 %(called \emph{gauge transformation} in Physics) 
    $h_v \mapsto h_v \tau_v$, $\sigma_v \mapsto \sigma_v \tau_v$, if $\tau$ is a
 codeword. As a matter of fact, the transformed HLLRs $h_v \tau_v$ are those 
 received when the codeword $\tau$ was transmitted, instead of the all-$+1$
 codeword.

 We now perform an average over all codewords $\tau$, obtaining
\begin{align*}
	\eep{h}{\angbra{\sigma_b}{}^{m}} = \frac{1}{|\loccode(G)|} \sum_{\tau\in
  \loccode(G)} \int \angbra{\sigma_b \tau_b}{}^{m} \prod_{v \in V} e^{h_v
  \tau_v} \d{\msf s}(h_v),
\end{align*}
where $\loccode(G)$ is the set of all codewords.

Note that the Gibbs bracket above averages over $\sigma$, and thus we can safely take $\tau_b$ out of the bracket. Since $m$ is odd, $\tau_b^m = \tau_b$. Next we use the definition of Gibbs
measure (equation \eqref{eq-gibbs}) to replace $\sum_{\tau \in \loccode(G)}
e^{\uln h \cdot (\uln \tau - \uln 1)} \tau_b$ with $e^{\sum_v h_v} Z(G) \angbra{\tau_b}{}$. We obtain
\begin{align}
	\eep{h}{\angbra{\sigma_b}{}^{m}} = 
  \frac{1}{|\loccode(G)|} \int e^{\sum_v h_v} Z(G)
  \angbra{\sigma_b}{}^{m+1} \prod_{v \in V} \d{\msf s}(h_v) .
\end{align}

Expanding $e^{\sum_v h_v} Z(G)$ into $\sum_{\lambda \in \loccode(G)} e^{h \cdot \lambda}$
we get
\begin{align*}
	\eep{h}{\angbra{\sigma_b}{}^{m}} = 
  \frac{1}{|\loccode(G)|} \!\! \sum_{\lambda \in \loccode(G)} \!\! \int 
  \angbra{\sigma_b}{}^{m+1} \!\!\prod_{v \in V} e^{h_v \lambda_v}\d{\msf s}(h_v)  .
\end{align*}

A second gauge transformation $h_v \mapsto h_v \lambda_v$, $\sigma_v \mapsto
\sigma_v \lambda_v$ allows us to cancel all
$\lambda$ factors, since $\lambda_v^2 = 1$. All 
$|\loccode(G)|$ terms in the sum are equal, so the expression simplifies to

\begin{align}
	\eep{h}{\angbra{\sigma_b}{}^{m}} = 
    \int \angbra{\sigma_b}{}^{m+1}\prod_{v \in V}  e^{h_v }\d{\msf s}(h_v)   ,
\end{align}
and thus the claim follows.
\end{IEEEproof}

The next result quantifies the effect on $\log Z$ of one extra check node added to some general linear code. This is the main reason why we chose to work with $\log Z$ instead of the conditional entropy.

\begin{lemma} \label{lem-z-expl}
	Given any graph $G$ and an additional check constraint $b$, we have that
	\[
        \eep{h}{\log  Z(G \cup b) - \log Z(G)} = -\log 2 + \sum_{r \in 2\zz_+} \frac{\eep{h}{\angbra{\sigma_b}{G}^{r}}}{r^2 - r}.
	\]
    In particular, $ -\log 2 \leq \log Z(G \cup b) - \log Z(G) \leq 0. $
\end{lemma}
The second part of the statement shows that the contribution of one extra check node gives only a finite variation in $\log Z$, and it turns out to be very useful for the cases where we need to show that two similar ensembles have log-partition functions that are asymptotically identical. 
\begin{IEEEproof}
Using the definition of the partition function $Z(G \cup b)$, we are able to write
\begin{align*}
	Z(G \cup b) \! &= \sum_{\uln \sigma \in \bin^V}e^{\uln h \cdot (\uln \sigma - \uln 1)} \frac{1 + \sigma_b}{2}  \prod_{a \in G} \frac{1 + \sigma_a}{2} \\&= Z(G) \angbra{\frac{1 + \sigma_b}{2}}{G}\!\!.
	\end{align*}

    Then $\log Z(G \cup b) - \log Z(G) = - \log 2 + \log (1 + \angbra{\sigma_b}{})$. Expanding the logarithm into power series, we obtain
\begin{align}
	\log (1 + \angbra{\sigma_b}{}) = \sum_{j \geq 1} \frac{(-1)^{j+1}}{j} \angbra{\sigma_b}{}^j.
\end{align}
We now use the Nishimori Identities (Lemma \ref{lem-nishi}) with  
	$\eep{h}{\angbra{\sigma_b}{}^{j-1}} = \eep{h}{\angbra{\sigma_b}{}^{j}}$,
	for even $j$. 
	This allows us to merge each odd-index term with the following term, proving the claim.
\end{IEEEproof}

Let us now analyze the terms of the form $\angbra{\sigma_b}{G}^{r}$ that appear in the last
lemma. For this purpose, we will work with the product measure $\mu^{\otimes
r}$. The measure space here is the one of $r$-tuples $(\sigma^{(1)}, \ldots,
\sigma^{(r)})$, where $\sigma^{(j)} \in \bin^V$.
Because the product measure is just the measure of $r$ independent copies of
the measure (henceforth called \emph{replicas}), it is easy to check that
\[
	\angbra{\sigma_b}{G}^{r} = \sigrepgold^{\otimes r}.
	\]

The $\otimes r$ sign at the top right of the bracket is just to remind us that
we deal with the product measure $\mu^{\otimes r}$. Since this is evident from context, we will
drop this sign in the future. We are then able to restate the last lemma as
follows.

\begin{corollary} \label{cor-fin-var}
	Given any graph $G$ and an additional check constraint $b$, we have that
	\begin{align}
		&\eep{h}{\log Z(G \cup b) - \log Z(G)} = 
		\nonumber\\ &\quad\quad\quad= 
		-\log 2 + \eepnob{h}{\sum_{r
			\in2\zz_+}
		\frac{\sigrepgold}{r^2 - r}}.
	\end{align}
\end{corollary}

\section{The configuration model} \label{sec-conf-model}

In this section we introduce the language needed to describe and dissect all the kinds of ensembles that we need. This section contains the core of the argument, albeit in terms of the elementary parts of ensembles. These parts are then put together in the next section, using interpolation. For this reason, the purpose of the constructs introduced in this section may only become clear a posteriori.

We assume that the configuration pattern introduced in Section \ref{subsect-coupled} is already fixed, i.e., it has been properly sampled at an earlier stage, 
and there are at least $N\dbar(1 - N^{-\eta})$  and at most $N\dbar(1 + N^{-\eta})$ sockets at every
position of the chain. By a straightforward application of a Azuma-Hoeffding type of inequality and the union
bound for all positions, this happens with high probability\footnote{By \emph{with high probability}
    we mean that the event in question happens with probability $1 - o(1/\mathrm{poly}(N))$. The
        parameters $L$ and $W$ are considered constant for this purpose.} in the first stage, as long as $0 < \eta <\frac{1}{2}$.
The fixed underlying configuration pattern is always of the 
coupled kind, i.e., there are $L$ groups of $N$ variable nodes each; the simple kind will arise from the conditions $W=1$ and $W=L$. 

Given the fixed configuration pattern, each variable node $v$ has a \emph{target degree} $d(v)$,
and exactly $d(v)$ \emph{sockets} numbered from $1$ to $d(v)$. Given a socket
$s$, let $\varn(s)$ denote the variable node that it is part of; by $\sigma_s$ we understand $\sigma_{\varn(s)}$. Let $\pos(v)$ denote the position of the variable $v$, with the notation extending to sockets in the obvious manner: $\pos(s) =
\pos(\varn(s))$.
We also set $S$ to be the set of all sockets and put $S_z = \{ s \in S : \pos(s) =
z \}$, i.e. the set of sockets at a particular position.

Check nodes will connect to sockets, so a check node $a$ will have the
form of a $K$-tuple $(a_1, \ldots, a_K)$, where the components $a_j$ are sockets. Note that the ordering of the edges leaving the check-node matters, so the check also ``stores'' this information. We say that
a check node $a$ has \emph{type} $\alpha = (\alpha_1, \ldots, \alpha_K)$ if
$\alpha_j = \pos(a_j)$, for all $1 \leq j \leq K$. 
In other words, the type
records the positions of the variable nodes to which the check node $a$
connects. 

We now consider random types, of which there are three kinds that are important
to us:
\begin{itemize}
	\item {\bf The connected random type.} This random type is uniformly distributed over
		the set of all $L^K$ possible types. We denote this distribution by $\conn$.
	\item {\bf The disconnected random type.} This type is uniformly
		distributed over the set of all types whose entries are all equal, i.e., types of the form $(z,z,\ldots, z)$.
		We denote this distribution by $\disc$.
	\item {\bf The coupled random type.} We choose a position $z$ uniformly at
		random and the result is a type uniformly distributed over the set of all types
		whose entries lie in the set $\{z, \ldots z + W - 1\}$. We denote this
		distribution by $\coup$.
\end{itemize}

We now define the \emph{positional occupation vector} $\occ_\alpha$ of a type
$\alpha$ to be a vector whose $z$ entry counts the number of occurrences of
position $z$ in type $\alpha$. As an example, if $K = 6$ and $\alpha = (1, 3, 2, 5, 1, 3)$
and assuming there are $L = 5$ positions, then $\occ_\alpha = (2, 1, 2, 0, 1)$.

Given a multiset of types $\Gamma$ (a set of types where duplicates can appear), we extend the definition of the positional
occupation vector to $\occ_\Gamma = \sum_{\alpha \in \Gamma} \occ_\alpha$. 

We
call a multiset of types \emph{$m$-admissible} if $\occ_\Gamma(z) \leq |S_z|-m$, for all
positions $z$. In other words, an $m$-admissible set of types $\Gamma$ ensures that
there exists a graph $G$ whose check constraints match one-to-one the types in
$\Gamma$ (we say that $G$ is \emph{compatible} with $\Gamma$), and
in addition, there are at least $m$ sockets at each position that remain
free. We will also use the word \emph{admissible} to mean $0$-admissible. 
One should think about the multiset of types as being a kind of ``pre-graph'',
where only the positions of the edges are decided, but not yet the actual
sockets.

The random graph generated by an admissible multiset of types $\Gamma$ is simply
given by the uniform measure over all graphs that are compatible with $\Gamma$.
To sample this random graph, the algorithm is as follows: start with the empty
graph; for each type
$\alpha = (\alpha_1, \ldots, \alpha_K)$ in the
multiset $\Gamma$ (the order is immaterial), pick \emph{distinct} $a_i$ uniformly at random
from the free sockets at position $\alpha_i$, and add check constraint $(a_1,
\ldots, a_K)$ to the graph. We will use this check-generating procedure often,
so we will say that check constraint $a$ is chosen according to distribution
$\nu(\alpha, {G})$
that depends on the type $\alpha$, and the part
${G}$ of the graph that is already in place.
Let $B_\alpha$ be the set of check constraints that are compatible with
$\alpha$ and are connected to free sockets (sockets that do not appear in
${G}$). Note that a socket must never be used twice, so they are chosen
without replacement. Then $\nu(\alpha, {G})$ is the uniform measure on
$B_\alpha$. 

We also trivially extend this definition to the case of a random graph
generated by a \emph{random} multiset of types. This latter random object will
be typically a list of independent random types of one of the three kinds
\emph{connected}, \emph{disconnected} and \emph{coupled}.
For the sake of
precision, in case the multiset of types is not admissible (by this we mean $m$-admissible, where $m$ will be fixed later), we define the generated
random graph to be the empty one.

We now introduce a quantity inspired from statistical physics that plays an
important role in what comes next, namely the \emph{positional overlap functions}.
Fix a configuration graph $G$, a channel realization $h$, and the number $r$ of
replicas of the measure $\mu_{G,h}$. Let $F_z \subseteq S_z$ be the set of free sockets
at position $z$ (free sockets being those that do not appear in any check
constraint of $G$). The \emph{positional overlap functions} $Q_z$, indexed by a position
$z$, are defined by

\begin{align}
	Q_z(\sigma^{(1)}, \ldots, \sigma^{(r)}) = \frac{1}{|F_z|} \sum_{s \in F_z}
	\sigma^{(1)}_s \cdots \sigma^{(r)}_s.
\end{align}

The next statement describes the link between the overlap functions and the replica averages introduced by Lemma \ref{lem-z-expl}. 

\begin{lemma}\label{lem-core}
	Given a number $m > K^2$, a fixed channel realization, a fixed graph $G$ whose associated type set is $m$-admissible
	and fixed type $\alpha$, we have
	\begin{align} \label{eq-lem-core}
		&\eepnob{a : \nu(\alpha,G)}{\sigrepg} =
		\nonumber\\&\quad\quad= 
		\angbra{ \prod_{j = 1}^K
		Q_{\alpha_j}\!(\sigma^{(1)}, \ldots, \sigma^{(r)})}{} +
		O\left(\frac{1}{m}\right).
	\end{align}
\end{lemma}
\begin{IEEEproof}
	The left hand side is nothing
	else than the average over all possible $a$ that are compatible with the type
	$\alpha$ and connect to free sockets. In other words,
	\begin{align}\label{eq-avg-b}
		\frac{1}{|B_\alpha|} \sum_{a \in B_{\alpha}} \sigrep. 
	\end{align}
	The goal is to somehow factorize the sum, but the fact that sockets 
	are not replaced makes it a bit harder. Suppose that, contrary to our current model, free sockets are
	allowed to be chosen with replacement, that is, it is possible to have $a_i = a_j$ for
	$i \neq j$. Let $B_{\alpha}'$ be the set of all (pseudo-)check constraints that are
	compatible with $\alpha$, and where sockets are allowed to appear multiple
	times. Then $B_{\alpha}'$ can be written as a product:
	\[
		B_{\alpha}' = F_{\alpha_1} \times \ldots \times F_{\alpha_K},
		\]
	where the set $F_z$ is the set of free sockets at position $z$. The idea is
	now that we can replace $B_\alpha$ with $B_{\alpha}'$ in the average
	\eqref{eq-avg-b} without losing too much, while gaining the ability to
	factorize the sum.

	The relation between the two, which is proven in Appendix
	\ref{ap-proof-b}, is 
	\begin{align} \label{eq-approx-b}
		&\frac{1}{|B_\alpha|} \sum_{a \in B_{\alpha}} \sigrep =\nonumber\\
		&\quad\quad= 
		\frac{1}{|B'_\alpha|} \sum_{a \in
			B'_{\alpha}} \sigrep + O\left( \frac{1}{m}\right).
	\end{align}

	Now we are in a better position, since on the r.h.s. any entry $a_i$ 
is chosen independently of the others. We rewrite the sum over $B'_\alpha$ in
the following way:
\begin{align*}
	\frac{1}{|F_{\alpha_1}|} \!\sum_{a_1 \in F_{\alpha_1}} \!\!\!\cdots
	\frac{1}{|F_{\alpha_K}|} \!\sum_{a_K \in F_{\alpha_K}}
	\!\!\!\angbra{\sigma_{a_1}^{(1)} \cdots\sigma_{a_K}^{(1)} \cdots \sigma_{a_1}^{(r)} \cdots
	\sigma_{a_K}^{(r)}}{}.
\end{align*}

Taking the bracket outside and factorizing, we obtain
\begin{align*}
	\angbra{\!
	\Bigg(\frac{1}{|F_{\alpha_1}|} \!\!\sum_{a_1 \in F_{\alpha_1}} 
	\!\!\!\!\sigma_{a_1}^{(1)} \cdots \sigma_{a_1}^{(r)}\Bigg) 
	\!\!\cdots\!\!
	\Bigg(\frac{1}{|F_{\alpha_K}|} \!\!\sum_{a_K \in F_{\alpha_K}} 
	\!\!\!\!\sigma_{a_K}^{(1)} \cdots \sigma_{a_K}^{(r)}\Bigg) 
	\!}{}\!,
\end{align*}
which we can identify as the bracketed product of positional overlap functions
on the right hand side of \eqref{eq-lem-core}.
\end{IEEEproof}

\begin{lemma} \label{lem-one-check}
	Let $G$ be a graph whose type multiset is $m$-admissible, and fix the channel
	realization $h$. Then the following inequalities hold:
	\begin{align}
		&\eepnob{\substack{\alpha : \conn \\ a : \nu(\alpha,G)}}{\sigrepg}
		\leq 
		\nonumber\\&\quad\quad\leq
		\eepnob{\substack{\alpha : \coup \\ a : \nu(\alpha,G)}}{\sigrepg}
		+ O(1/m),\\
		&\eepnob{\substack{\alpha : \coup \\ a : \nu(\alpha,G)}}{\sigrepg}
		\leq \nonumber\\
		&\quad\quad\leq
		\eepnob{\substack{\alpha : \disc \\ a : \nu(\alpha,G)}}{\sigrepg}
		+ O(1/m).
	\end{align}
\end{lemma}
\begin{IEEEproof}
	The claim follows by Lemma \ref{lem-core} if we manage to show the
	following two inequalities:
	\begin{align} \label{eq-conv1}
		\eepnob{\alpha : \conn}{\angbra{Q_{\alpha_1} \cdots Q_{\alpha_K} }{}} 
		&\leq
		\eepnob{\alpha : \coup}{\angbra{Q_{\alpha_1} \cdots Q_{\alpha_K}}{}}, \\
		\eepnob{\alpha : \coup}{\angbra{Q_{\alpha_1} \cdots Q_{\alpha_K}}{}} 
		&\leq \label{eq-conv2}
		\eepnob{\alpha : \disc}{\angbra{Q_{\alpha_1} \cdots Q_{\alpha_K}}{}},
	\end{align}
	where the dependence of the positional overlap functions on the spin
	systems $\sigma^{(j)}$ has
	been dropped in order to lighten notation.
	
%	 \eqref{eq-conv1}, we  
	We rewrite the quantities above as follows:
	 \begin{align}
		%conn
		 \label{eq-fin-conn}
		 &\eepnob{\alpha : \conn}{\angbra{Q_{\alpha_1} \cdots Q_{\alpha_K} }{}}
		  = \nonumber\\&\quad =
		 \frac{1}{L^K} \!\! \sum_{\substack{(\alpha_1, \ldots, \alpha_K)\\ \in
		 [L]^K}} \!\!\!  \angbra{Q_{\alpha_1} \cdots Q_{\alpha_K}}{} 
		 \!=\! \angbra{\!\Bigg(\! \frac{1}{L} \!\sum_{z \in [L]} \!Q_z\!\Bigg)^{\!\!K}}{}, \\
		%coup
		 &\eepnob{\alpha : \coup}{\angbra{Q_{\alpha_1} \cdots Q_{\alpha_K} }{}} 
		  = \nonumber\\&\quad =
		 \frac{1}{L} \sum_{z' \in [L]} \frac{1}{W^K} \mkern-15mu \sum_{\substack{(\alpha_1, \ldots, \alpha_K)\\ \in
			 \{z', \ldots, z'+W-1\}^K}} \mkern-27mu  \angbra{Q_{\alpha_1} \cdots
			 Q_{\alpha_K}}{}\nonumber\\
		 &\quad= \angbra{\frac{1}{L} \sum_{z' \in [L]} 
		 \Bigg( \frac{1}{W} \sum_{z  = z'}^{z'+W-1}
		    Q_z\Bigg)^{\!\!K}}{}, \\
		% disc
		\label{eq-fin-disc}
		 &\eepnob{\alpha : \disc}{\angbra{Q_{\alpha_1} \cdots Q_{\alpha_K} }{}} 
		  = \nonumber\\&\quad =
		 \frac{1}{L} \sum_{z \in [L]} \angbra{Q_{z} \cdots Q_{z}}{} 
		 = \angbra{\frac{1}{L} \sum_{z \in [L]} 
		 	Q_z^K}{}.
	 \end{align}
     In the above expressions we assume $Q_z$ is defined for all integer $z$ using the relation $Q_{z'} = Q_{z''}$ whenever $z' \equiv z'' (\mod L)$.
	 Both inequalities \eqref{eq-conv1} and \eqref{eq-conv2} are proved by an
	 application of Jensen's Inequality using the convexity of the
	 function ${x\mapsto x^K}$, for even $K$.
\end{IEEEproof}

\section{The interpolation} \label{sec-interp}

We now move a bit further and consider random ensembles of graphs. 
These are
obtained in the following way: first we prescribe the numbers of random types of
each kind that we want, i.e. how many types should be connected, disconnected
and coupled. Afterwards, the random types are sampled according to the
distributions
prescribed. Finally the graph is chosen uniformly to match the multiset of
types, in the spirit of the previous section. 

We use the notation $G : \left\{\substack{t_1 \times \coup \\ t_2 \times
\disc}\right\}$ to say that $G$ is sampled in the way outlined above, where
$t_1$ and $t_2$ are the number of random types of the coupled kind and
disconnected kind, respectively. Of course, we could specify any combination of
the three kinds, $\conn$ included.

Now we need to set the number of check nodes in the ensemble. There are two
conflicting constraints we would like to satisfy: first, the set of types needs to be
admissible with high probability --- so that the sampled graph exists in the
form we want; second, the number of free sockets that remain should be small, in the sense that the proportion of free sockets needs to vanish in the limit. 

The average amount of check nodes needed to use all available sockets is (ideally) $N L \dbar / K$. 
However, there is a fluctuation ($\pm N^{1-\eta} \dbar$ at each position) of the amount of available sockets and it might not be possible to connect actual check nodes to all sockets (for example, because of window constraints). 
As a consequence, we choose the actual size of the graph (by this we mean the number of multi-edges, i.e. check nodes) to be $T = N L \dbar (1 - N^{-\gamma}) / K$,
so in case the graph is admissible there will be $O(N^{1-\gamma})$ free
sockets left at each position. The exponent $\gamma$ is arbitrary, as long as $0 < \gamma < \eta$. The next lemma confirms that by using this
value for $T$, the resulting set of types is admissible with high probability.
% TODO: add the $L$ in the O notation.
\begin{lemma} \label{lem-hi-prob}
	Let $\alpha^{1}, \ldots, \alpha^{T}$ be random types, each drawn from
	a distribution that is either $\conn$, $\disc$ or $\coup$ (could be different for each type). Then with high
	probability (more precisely $1 - O(\exp(-\kappa N^{1 - 2\gamma}))$, for some positive constant $\kappa$) the resulting
	multiset of types is $\dbar N^{1-\gamma}/2$-admissible.%, for any $\delta < \gamma$.
	%as long as $\gamma > 1/2$.
\end{lemma}
\begin{IEEEproof}
	The plan is the following: fix a position $z$, and show that the number of
	appearances of $z$ as entries of $\alpha^{1}, \ldots, \alpha^{T}$
	exceeds $TK/L + \dbar N^{1-\gamma}/2$ with a very small probability. Next, by the
	union bound over all positions $z$, we upper bound the
	probability that the graph is not $\dbar N^{1-\gamma}/2$-admissible and the lemma
	is proved.

	We concentrate on the above claim, and define $X_t$ to be the number of
	entries in $\alpha^t$ equal to $z$, for $1 \leq t \leq T$. Clearly the $X_t$ are independent, bounded
	and their expectation equals $K/L$ (the choice of distribution of $\alpha^t$ is
	immaterial as long as it is one of $\conn$, $\disc$ or $\coup$). 
	Then by Hoeffding's Inequality, the probability that $\sum X_t$ deviates from
	its expectation $TK/L$ decays very fast. More exactly,
	\begin{align}
		&\pr{\sum_{t = 1}^T X_t \geq \frac{TK}{L} + \frac{1}{2} \dbar N^{1 - \gamma}} 
		\!\leq 
	%	\nonumber\\&\quad\quad\leq 
		\exp\!\left(-\frac{\dbar^2 N^{2 - 2\gamma}}{2K^2 T}\right),
	\end{align}
	which proves the claim.
\end{IEEEproof}

The previous lemma essentially allows us to take the expectation over an
ensemble of graphs without caring too much about non-admissibility. 
This enables us to prove the following key lemma.

\begin{lemma} \label{lem-ineq}
	The following two inequalities hold:
	\begin{align} \label{eq-end-chain}
		&\eepnob{h, G : \left\{T \times \conn\right\}}{\log Z(G)} 
		\leq\nonumber\\&\quad\quad\leq
		\eepnob{h,G : \left\{T \times \coup\right\}}{\log Z(G)} + O\left(N^{\gamma}\right), \\ 
		&\eepnob{h, G : \left\{T \times \coup\right\}}{\log Z(G)} 
		\leq\nonumber\\&\quad\quad\leq
		\eepnob{h,G : \left\{T \times \disc\right\}}{\log Z(G)} + O\left(N^{\gamma}\right). 
	\end{align}
\end{lemma}
\begin{IEEEproof}
	We only discuss the first of the two inequalities, since the proof of the other is identical. We will set up a chain of inequalities, at the ends of which sit the two
	quantities that we need to compare. This is the main idea of the \emph{interpolation method}: finding a sequence of objects that transition ``smoothly'' between two
	objects that can differ significantly. In our case, it is easily seen that the claim follows if we are able to show that
	\begin{align} \label{eq-chain-spot}
		&\eepnob{h, G : \left\{\substack{(t+1) \times \conn \\ (T-t-1) \times \coup}\right\}}{\log Z(G)} \leq\nonumber\\          
		  &\quad\quad\leq 
		  \eepnob{h, G : \left\{\substack{t \times \conn \\ (T-t) \times \coup}\right\}}{\log Z(G)}+ O\left(N^{\gamma - 1}\right).
	\end{align}
	The two ensembles involved in inequality \eqref{eq-end-chain} lie at the endpoints of a chain of $T$ inequalities of the form above, with $t$ moving from $0$ to $T-1$.
	The crucial observation here is that the two ensembles $\left\{\substack{(t+1) \times \conn \\ (T-t-1) \times \coup}\right\}$ and $\left\{\substack{t \times \conn \\
	(T-t) \times \coup}\right\}$ can both be obtained by sampling a graph $\widetilde{G}$ from their common part, $\left\{\substack{t \times \conn \\ (T-t-1) \times
	\coup}\right\}$ and in case $G$ is not null, adding an extra random check constraint sampled according to $\conn$ and $\coup$, respectively. The plan is to show that
	the inequality \eqref{eq-chain-spot} holds also when $\widetilde{G}$ is fixed, and then to average over $\widetilde{G}$.

	Let us fix $m = \dbar N^{1 - \gamma}/2$, and let us first deal with the case when the realization of the ensemble $\left\{\substack{t \times \conn \\
	(T-t-1) \times \coup}\right\}$ is not $m$-admissible. This event 
	occurs with a very small probability, subexponential according to Lemma \ref{lem-hi-prob}.
	Since $\log Z(G) = O(N)$ (according to Lemma \ref{lem-z-expl}), the error obtained by not considering this case is extremely small 
	and fits in the tolerated
	term $O\left(\frac{1}{N^{1-\gamma}}\right)$.

	Otherwise, $\widetilde{G}$ is such that there are at least $m$ free sockets at every position, and we need to show that
	\begin{align*}	
		\eesym_h\eepnob{\substack{\alpha : \conn \\ a : \nu(\alpha,\widetilde{G})}}{\log Z(\widetilde{G} \cup a)} &\leq\eesym_h\eepnob{\substack{\alpha : \coup \\ a : \nu(\alpha,\widetilde{G})}}{\log
			Z(\widetilde{G} \cup a)}.
	\end{align*}	
	
	We subtract $\log Z(\widetilde{G})$ on both sides and then use Lemma \ref{lem-z-expl} to write the difference of log partition functions as a linear combination of
	brackets of the form $\angbra{\sigma^{(1)}\cdots\sigma^{(r)}}{\widetilde{G}}$, after which we can readily apply Lemma \ref{lem-one-check} and the claim follows.
\end{IEEEproof}

%We next
%compare ensembles that differ in the kind (by kind we mean 
%$\conn$, $\disc$ or $\coup$) of exactly one of their component random
%types.

%\begin{align}
%	\eep{G : \left\{\substack{n \times \coup \\ (m-n) \times \disc}\right\}}{mc^2}
%\end{align}

\section{Retrieving the original LDPC ensembles} \label{cod:sec6}

We will now investigate further the connection between the ensembles $\left\{T \times \conn\right\}$ and $\left\{T \times \disc\right\}$. In fact, they are both variants of the uncoupled ensembles introduced in the beginning of Section \ref{sec-prelim}. The first one is very similar to $\ldpc(NL,\Lambda,K)$, and the second one is similar to $L$ copies of $\ldpc(N,\Lambda,K)$. The only differences that occur are related to the case where there is a large deviation in the number of sockets generated in the first stage, or when the multisets of types generated by $\left\{T \times \conn\right\}$ and $\left\{T \times \disc\right\}$ are not admissible. Also since the first stage of the ensemble generation, where we obtain the configuration pattern, is the same in all cases, we condition on the event that the configuration pattern is known and that it satisfies the condition stated at the beginning of Section \ref{sec-conf-model}, namely that the number of sockets at each position is $N\dbar / K \pm O(N^\eta)$. 

We can easily see that the ensemble $\left\{T \times \disc\right\}$, conditioned on the fact that its realization is admissible, can be extended to $L$ copies of
the simple (i.e. uncoupled) ensemble on $N$ variable nodes by adding $O(N^{1 - \gamma})$ extra check constraints. Thus the scaled log partition function is the same up to a sub-linear term.

Can we say the same about the ensemble $\left\{T \times \conn\right\}$ and the simple ensemble on $NL$ variable nodes? Yes, but it requires a lengthier argument. Let us
look closer at the latter. This ensemble is not generated using types (since positions play no role here), but we can still count the occurrences of various types that appear in it. There are exactly $L^K$
different types, and the next proposition estimates the probability that a particular random check constraint in the simple ensemble $\ldpc(NL,\Lambda,K)$ has a certain type. To see the crux of the problem, in the $\left\{T \times \conn\right\}$ ensemble, the types are generated uniformly. Whereas in the simple ensemble, a position with considerably more occupied sockets than other positions has a lesser chance to be picked.

We will proceed by transforming the ensemble $\ldpc(NL,\Lambda,K)$ (the \emph{simple} ensemble) into $\left\{T \times \conn\right\}$ (the \emph{connected} ensemble) through only a small amount of check additions and deletions. Let $\oldx_\alpha$ be the number of check nodes of type $\alpha$ that occur in a realization of the simple ensemble. For every type $\alpha$, let $\oldy_\alpha$ be a random variable sampled according to $\mathrm{Bin}(T, L^{-K})$. If $\oldx_\alpha > \oldy_\alpha$, then exactly $\oldx_\alpha - \oldy_\alpha$ check nodes of type $\alpha$ selected uniformly at random from the existing ones are deleted from the simple ensemble. Otherwise, exactly $\oldy_\alpha - \oldx_\alpha$ check nodes of type $\alpha$ are chosen uniformly at random from all possible combinations of compatible free sockets and inserted in the graph without replacement. All insertions of check nodes must occur after all deletions have been performed (the order of the types is important). If at any stage there are no free sockets at a particular position 
to choose from, it just means the underlying multiset of types (which here is given by the numbers $\oldy_\alpha$) is not T-admissible, and we produce the trivial code. 

In order to bound the number of check node insertions and deletions, we compute the first and second moments of $\oldx_\alpha - \oldy_\alpha$. The total number of check nodes $M$ in the simple ensemble is fixed for our purposes (depends only on the configuration pattern), so we can write $\oldx_\alpha = \sum_a R^a_\alpha$, where $R^a_\alpha$ is the indicator random variable of the event that check node $a$ has type $\alpha$, and the sum ranges over all $M$ check nodes.

\begin{proposition}
	The expectation and variance of $\oldx_\alpha - \oldy_\alpha$ are given by
	\begin{align}
		\ee{\oldx_\alpha - \oldy_\alpha} &= O(N^{1-\gamma}), \\
		\mathrm{Var} [\oldx_\alpha - \oldy_\alpha] &= O(N^{2-\eta}).
	\end{align}
\end{proposition}
\begin{IEEEproof}
	We determine first the probability $\eenob{R^a_\alpha}$ that a check node $a$ has type $\alpha$. This event happens if and only if all sockets	$a_i$ to which $a$ is connected are placed at positions $\alpha_i$. For this, we need to evaluate the proportion of free sockets at each position (all sockets are free initially, because w.l.o.g. we can say that $a$ is the first check node to be allocated). The number of sockets at any position is between $N\dbar(1 - N^{-\eta})$ and $N\dbar(1 + N^{-\eta})$; the number of occupied sockets is at most $K-1$ (from previous edges). Thus, the probability that $\pos(a_i) = \alpha_i$ is lower-bounded by 
	\[\frac{N\dbar(1 - N^{-\eta}) - K}{NL\dbar(1 + N^{-\eta})} = \frac{1}{L} - O(N^{-\eta}),\]
and, likewise, upper-bounded by
	\[\frac{N\dbar(1 + N^{-\eta})}{NL\dbar(1 - N^{-\eta})} = \frac{1}{L} + O(N^{-\eta}).\]
	It then follows that
\begin{align}
	\eenob{R^a_\alpha} = \left(\frac{1}{L} + O(N^{-\eta}) \right)^K = \frac{1}{L^K} + O(N^{-\eta}).
\end{align}

	For the second moments we need $\ee{R^a_\alpha R^b_\beta}$, i.e. the probability that $a$ and $b$ have types $\alpha$ and $\beta$ at the same time. The reasoning is essentially similar to the previous case, only now there are $2K$ edges to connect and at most $2K-1$ occupied sockets (by symmetry we can arrange that $a$ and $b$ are the first two check nodes to be allocated). Then we have

\begin{align}
	\ee{R^a_\alpha R^b_\beta} = \left(\frac{1}{L} + O(N^{-\eta}) \right)^{2K} \!\!\! = \frac{1}{L^{2K}} + O(N^{-\eta}).
\end{align}
	
	By summing over all check nodes, we get $\eenob{\oldx_\alpha} = \frac{M}{L^K} + O(N^{1-\eta})$ and after elementary calculations, $\mathrm{Var} \oldx_{\alpha} = O(N^{2-\eta})$.
	Since $\oldy_\alpha$ is binomially distributed, and using $T = M + O(N^{1-\gamma})$, we have 
	\[\eenob{\oldy_\alpha} = \frac{T}{L^K} = \frac{M}{L^K} + O(N^{1-\gamma}), \]
	and also \[\mathrm{Var} \oldy_\alpha = T \frac{1}{L^K} \left( 1 - \frac{1}{L^K} \right) = O(N),\]
	which is much smaller than $\mathrm{Var} \oldx_\alpha$.
\end{IEEEproof}

To show that the amount of inserted and deleted check nodes is small, we employ now the Chebyshev Inequality, which, for some value of the parameter $\zeta$ to be fixed shortly, reads  
\begin{align*}
	\pr{\left|\oldx_\alpha - \oldy_\alpha - O\left(N^{1-\gamma}\right)\right| \geq N^\zeta O\left(N^{1-\frac{\eta}{2}}\right)}
	     \leq \frac{1}{N^{2\zeta}}.
\end{align*}

We fix the values $\zeta = \frac{\eta}{4}$ and $\gamma = \frac{\eta}{2}$ (these choices are somewhat arbitrary), and simplifying we obtain
\begin{align*}
	\pr{\left|\oldx_\alpha - \oldy_\alpha\right| \geq O\left(N^{1-\frac{\eta}{4}}\right)}
	\leq N^{-\frac{\eta}{2}}.
\end{align*}

Using the union bound over all $L^K$ possible types, the bound on the probability that the number of insertions and deletions is sub-linear in the way depicted above remains $O\left(N^{-\eta /2}\right)$. In case the the number of insertions and deletions is too large, we use the $O(N)$ we use the fact that $\log Z(G)$ is always $O(N)$ (see Lemma \ref{lem-z-expl}). This proves the following lemma. 

\begin{lemma}
	Transmitting over a BMS channel, we have
	\begin{align*}
		&\eepnob{h,G:\ldpc(NL,\Lambda,K)}{\log Z(G)} \\&\quad\quad \geq \eepnob{h,G : \left\{T \times \conn\right\}}{\log Z(G)} + O\left(N^{1-\frac{\eta}{4}}\right).
	\end{align*}
\end{lemma}

\section{The large $N$ limit}
		\label{sec-large-limit}

	This section wraps up the proof of Theorem \ref{main-thm}. The main ingredient is the content of Lemma \ref{lem-ineq}, which can be written as 
\begin{align}
	&\eepnob{h, G : \left\{T \times \conn\right\}}{\log Z(G)} - O\left(N^{1-\gamma}\right)
	\nonumber\\&\quad\quad \leq 
	  \eepnob{h,G : \left\{T \times \coup\right\}}{\log Z(G)}
	\nonumber\\&\quad\quad \leq 
	  \eepnob{h,G : \left\{T \times \disc\right\}}{\log Z(G)} + O\left(N^{1-\gamma}\right). 
\end{align}

Using the results from the previous section on the comparison with the simple ensembles and scaling everything by $NL$, we obtain
\begin{align}
	&\frac{1}{NL} \eepnob{h, G : \ldpc(NL,\Lambda,K)}{\log Z(G)} - O\left(N^{-\gamma}\right) %\leq
	\nonumber\\&\quad\quad \leq 
	\frac{1}{NL} \eepnob{h,G : \left\{T \times \coup\right\}}{\log Z(G)}% \leq 
	\nonumber\\&\quad\quad \leq 
	\frac{1}{N} \eepnob{h,G : \ldpc(N,\Lambda,K) }{\log Z(G)} + O\left(N^{-\gamma}\right). 
\end{align}

The next step is to take the $N \to\infty$ limit, and in case it exists for the outer terms, which we are about to show, we can apply the ``sandwich rule'' to obtain Theorem \ref{main-thm}. 
Note that the ensemble appearing in the middle is what we call $\ldpc(N,L,W,\Lambda,K)$ --- we are of course not obliged to pick it as such: we could do another level of processing in the style of the previous section; however the current form is known to fulfill the Maxwell conjecture, so we need not go any further.

To show that the limit
\[\lim_{N\to\infty} \frac{1}{N}  \eepnob{h,G : \ldpc(N,\Lambda,K) }{\log Z(G)}\]
exists, we use the following result, whose proof can be found in the Appendix of \cite{bayati-gamarnik-tetali}.

\begin{lemma}[The modified superadditivity theorem]
	Given $\alpha \in (0,1)$, suppose a non-negative sequence $\{a_{N, N\geq 1}\}$ satisfies
\begin{align}
	a_{N_1 + N_2} \geq a_{N_1} + a_{N_2} - O(\left( N_1 + N_2 \right)^\alpha)
\end{align}
for every $N_1, N_2 \geq 1$. Then the limit $\lim_{N \rightarrow \infty} \frac{a_N}{N}$ exists (it may be $+\infty$).
\end{lemma}
The claim then follows by setting the sequence $a_N$ to be the negative of the sequence we study (since $\log Z(G)$ are negative). It remains to be shown that superadditivity indeed holds.

Since this part is a somewhat simpler variation of the interpolation we have already seen, we only present the proof sketch. We consider a coupled ensemble consisting of only two positions($L = 2$) and interpolate between the cases $W = 1$ (disconnected case) and $W = 2$ (connected case). The novelty is that the number of variables at the first and second positions differ, they are $N_1$ and $N_2$, respectively. For the connected case, when edges from check nodes are connected, we do not pick the position at random, but rather weigh the choice by $\nu_1 = \frac{N_1}{N_1 + N_2}$ and $\nu_2 = \frac{N_2}{N_1 + N_2}$, respectively. 

The only difference appears in the reasoning of Lemma \ref{lem-one-check}, where the types are not uniformly distributed anymore. The types are now binary strings of length $K$, with the two symbols appearing denoting the position, one having weight $\nu_1$, the other $\nu_2$. The weight of the type is the product of the weights of the symbols it contains. If $\alpha$ is a type, let $\nu(\alpha)$ be the weight of that type. Then Equations \eqref{eq-fin-conn} and \eqref{eq-fin-disc} become

 \begin{align*}
		%conn
		 &\eepnob{\alpha : \conn}{\angbra{Q_{\alpha_1} \cdots Q_{\alpha_K} }{}}
		  = \nonumber\\&\quad =
		 \!\!\sum_{\alpha \in \{1,2\}^K}\!\!\! \nu(\alpha)  \angbra{Q_{\alpha_1} \cdots Q_{\alpha_K}}{} 
			 = \angbra{(\nu_1 Q_1 + \nu_2 Q_2)^K}{}, \\
		% disc
		 &\eepnob{\alpha : \disc}{\angbra{Q_{\alpha_1} \cdots Q_{\alpha_K} }{}} 
		  = \nonumber\\&\quad =
		  \sum_{z \in \{1,2\}} \nu_z \angbra{Q_{\alpha_1} \cdots Q_{\alpha_K}}{} 
		  = \angbra{
		   \nu_1 Q_1^K + \nu_2 Q_2^K}{},
	 \end{align*}
and clearly the lemma remains true in this case as well.

\section{Proof of Theorem \ref{th-6}}\label{app-b}

We construct a smooth family of channels by interpolating
between the given channel $\msf c^*$ and the worst channel, denoted
by $\Delta_0$ (since in the log-likelihood representation it consists of
a point mass at 0):
   \[
   \msf c_\hc = \frac{\hc - \hc^*}{1 - \hc^*} \Delta_0 + \frac{1- \hc}{1 - \hc^*} \msf c^*,
   \]
where $\hc^* = H(\msf c^*)$ and the parameter $\hc$ has been chosen in such a way that it coincides
with $H(\msf c)$, varying from $\hc^*$ to $1$. Also, to ease notation, for the DE fixpoint we use
$\msf x_\hc$ as  a shorthand for $\msf x_{\msf c_\hc}$.

The plan is as follows: first we will show that
\begin{align} \label{eq-48}
 \frac{d}{d\hc} \Phi(\msf x_\hc, \msf c_\hc) = g^{\textrm{BP}}(\hc).
\end{align}

Then by Theorem \ref{th-4}, we can replace $g^{\textrm{BP}}(\hc)$ with $g^{\textrm{MAP}}(\hc)$. We 
integrate the two sides between $\hc^*$ and $1$ and check that for the worst channel
\[
\Phi(\msf x_1, \Delta_0) = R = \lim_{N\to\infty} \frac{1}{N} \eesym H(\underline{X}|\underline{Y}(1)),
    \]
thereby ending the proof of Theorem \ref{th-6}.

It remains to check \eqref{eq-48}. Note that an equivalent form of \eqref{eq-def-gbp} written in the
density evolution language is
\begin{align}\label{eq-49}
g^{\textrm{BP}}(\hc) = \left[ \frac{d}{d\hc} H(\msf c_\hc \opc \Lambda^\opc (\rho^\ops(\msf
                x_{\hc'}))) \right]_{\hc' = \hc}.
\end{align}

In the ensuing calculations, we will replace $\msf x_\hc$
by $\msf x$ whenever
its meaning is clear from context. It can be easily checked that
this form is very similar to the left hand side of \eqref{eq-48}, except
that the differential operator only affects $\msf c$ and not $\msf x$ (i.e. it
is a partial derivative). We will subsequently show that since
$\msf x$ is the forward DE fixpoint, the partial derivative equals the
total derivative.

\newcommand{\xx}{\msf x}
\newcommand{\yy}{\msf y}
\renewcommand{\zz}{\msf z}
\newcommand{\xxh}{\xx_\hc}
\newcommand{\cc}{\msf c}
\newcommand{\cch}{\cc_\hc}
\newcommand{\dxx}{\Delta\xx}
\newcommand{\dcc}{\Delta\cc}
\newcommand{\dhc}{\Delta\hc}

We will compute the derivative of each term in \eqref{eq-15} separately.
The treatment is somewhat similar to the calculation of
directional derivatives of the potential function in \cite{pfister-nicolas-kumar-young}. Each
of the first three terms is of the form
\begin{align*}
\frac{d}{d\hc} H(f^{\ops}(\xxh)) &= \lim_{\dhc\to 0} \frac{H(f^\ops(\xx_{\hc + \dhc})) -
    H(f^\ops(\xxh))}{\dhc} \\
        &= \lim_{\dhc\to 0} \frac{H(f^\ops(\xx + \dxx)) - H(f^\ops(\xx))}{\dhc},
\end{align*}
where $f(u) = \sum_k f_k u^k$ is some polynomial and $\dxx$ is a shorthand for $\xx_{\hc + \dhc} -
\xx_\hc$. To keep the formulas uncluttered, in all expressions containing the limit $\dxx \to 0$ we
suppress the $\hc$ indices. Expanding, we obtain
\begin{align*}
&\frac{d}{d\hc}H(f^{\ops}(\xxh)) = \lim_{\dhc\to 0} \frac{H\!\left( \sum_k f_k \sum_{j\geq 1}
        {k \choose j} \dxx^{\ops j} \ops \xx^{k-j} \right)}{\dhc} \\
    &= \!\lim_{\dhc\to 0}\!\!\frac{H\!\left( \sum_{k} k f_k \dxx \ops \xx^{k-1} \right)}{\dhc} \!+\!\!
        \lim_{\dhc\to 0}\!\!\frac{H\!\left(\dxx^{\ops 2} \ops g(\xx, \dxx) \right)}{\dhc} \\
    &= \!\lim_{\dhc\to 0}\!\frac{H\!\left( \dxx \ops f'^{\ops}(\xx) \right)}{\dhc},
\end{align*}
where in the last step all the higher order terms (i.e. those containing a $\opc$-power of $\dxx$
    higher than $1$ disappear.
The
polynomial $g$ was introduced just to collect those terms, and
the fact that they vanish is shown below in Lemma \ref{lem-24}.
Explicitly, the derivatives of the first three terms are:
\begin{align*}
\frac{d}{d\hc} \left[ -\frac{\Lambda'(1)}{P'(1)} H(P^\ops(\xxh))  \right] \!&=\! - \Lambda'(1)
\!\!\lim_{\dhc\to 0}\!\!  \frac{H\left( \dxx \ops \rho^{\ops}(\xx)\right)}{\dhc}, \\
\frac{d}{d\hc} \left[ -{\Lambda'(1)} H(\rho^\ops(\xxh))  \right] \!&=\! - \Lambda'(1)
\!\!\lim_{\dhc\to 0}\!\! \frac{H\left( \dxx \ops \rho'^{\ops}(\xx)\right)}{\dhc}, \\
\frac{d}{d\hc} \left[ {\Lambda'(1)} H(\xxh \ops \rho^\ops(\xxh))  \right] \!&=\! \\
 &\mkern-180mu =\Lambda'(1) \lim_{\dhc\to 0} \frac{H\left( \dxx \ops \rho^{\ops}(\xx)\right)
    + H\left( \dxx \ops \xx \ops \rho'^{\ops}(\xx)\right)}{\dhc}.
\end{align*}

Using Lemma \ref{lem-22}, we replace $ H\left( \xx \ops \rho'^{\ops}(\xx)\ops \dxx \right)$ with 
$ H\left( \rho'^{\ops}(\xx) \ops \dxx \right) -H\left( \xx \opc (\rho'^{\ops}(\xx) \ops \dxx)
    \right)$, and we are thus able to cancel the contributions of the first two terms.

The derivative of the last of the four terms in \eqref{eq-15} needs to be handled more carefully,
    since it contains both kinds of operations on densities. However, the idea remains the same: we
    examine the quantity

    \[
    H(( \cc + \dcc) \opc \Lambda^\opc(\rho^\ops(\xx + \dxx)) - \cc \opc
            \Lambda^\opc(\rho^\ops(\xx)))
    \]
   andi we classify the terms that appear according to the position
of $\dcc$ and $\dxx$. There are two terms that contain once either
$\dcc$ and $\dxx$:
\begin{itemize}
\item $\dcc \opc \Lambda^\opc(\rho^\ops(\xx)),$
\item $\cc \opc \Lambda'^\opc(\rho^\ops(\xx)) \opc (\rho'^\ops(\xx) \ops \dxx).$
\end{itemize}
The higher order terms (the ones that contain at least two of $\dxx$ and $\dcc$) are of the types
\begin{itemize}
\item $(\dxx \ops \dxx \ops g_1(\xx, \dxx)) \opc g_2(\xx,\dxx,\cc),$
\item $(\dxx \ops g_1(\xx)) \opc (\dxx \ops g_2(\xx)) \opc g_2(\xx,\dxx,\cc),$
\item $(\dxx \ops g_1(\xx,\dxx)) \opc g_2(\xx, \dxx) \opc \dcc,$
\end{itemize}
where the functions $g_1,g_2, g_3$ are products involving $\opc$ and $\ops$ of their parameters.
All the terms above have vanishing contributions in the limit, by Lemma \ref{lem-24}.

We are now able to collect all the terms that remain and
assemble them in the form
\begin{align*}
&\frac{d}{d\hc} U(\xxh, \cch) = \\
&\quad \lim_{\dhc\to 0}  \frac{H((\xx - \cc \opc \Lambda'^\opc(\rho^\ops(\xx))) \opc
        (\rho'^\ops(\xx) \ops \dxx))}{\dhc} \\
&\qquad +  \lim_{\dhc\to 0}  \frac{H(\dcc \opc
        \Lambda^\opc(\rho^\ops(\xx)))}{\dhc} \\
&\quad = 0 + g^{\textrm{BP}}(\hc),
\end{align*}
where in the last step we used the fact that $\xx$ is the fixpoint
of the DE equation, and also the alternative definition of the
BP GEXIT curve provided by \eqref{eq-49}.

The proof is now complete, and we are left to show that the
higher order terms do not contribute in the limit.We begin with
some definitions and some new notations. Degradation induces
a partial ordering on $\xspace$, which we denote by $z \prec z'$, where
$z'$ is degraded with respect to $z$. Note that density evolution
preserves degradation, and the following proposition follows
from standard arguments in \cite{ruediger-tom-book}.

\begin{proposition}
If $\cc,\cc' \in \xspace$ and $\cc \prec \cc'$ then $\xx_\cc \prec \xx_{\cc'}$.
\end{proposition}

There is a metric defined on $\xspace$, the \emph{Wasserstein distance (on the $|D|$ domain)}
\cite{shrini-tom-ruedi-universal}, that has the following useful properties which we state here
without proof. For any $\zz, \zz' \yy \in \xspace$,
        \begin{align*}
            d(\zz \opc \yy , \zz' \opc \yy) \leq 2d(\zz,\zz'),\\
            d(\zz \ops \yy , \zz' \ops \yy) \leq d(\zz,\zz').\\
        \end{align*}
Let $\mathcal{F}$ be the set of functions $f:\xspace\to\xspace$ of the form
\[
f(\zz) = \yy_1 *_1 (\yy_2 *_2 (\ldots (\yy_k *_k \zz)))
    \]
for some $\yy_1,\ldots, \yy_k \in\xspace$ and $*_1, \ldots, *_k \in \{\opc,\ops\}$. We can easily
extend $f$ by linearity, in order to define quantities like $f(\zz - \zz')$. Then for each $f \in
\mathcal{F}$ there is a constant $M$ such that for all $\zz \prec \zz'$ we have that
\begin{align} \label{eq-51}
\d(f(\zz), f(\zz')) \leq Md(\zz, \zz').
\end{align}

If $\zz \prec \zz'$, the Wasserstein distance is bounded above and below by powers of the
Bhattacharyya functional, in the sense that
\[
    \frac{1}{4}(\mathcal{B}(\zz') - \mathcal{B}(\zz))^2 \leq d(\zz,\zz') \leq
    2\sqrt{\mathcal{B}(\zz') - \mathcal{B}(\zz)}.
\]

The following lemma (part of Lemma 21 in \cite{shrini-tom-ruedi-universal}) will enable
us to factorize the entropy of a $\opc$-product. The reason why
we consider the Bhattacharyya functional is contained in the
following lemmas.

\begin{lemma}\label{lem-20}
Let $\zz,\zz',\yy,\yy' \in\xspace$ such that $\zz \succ \zz'$. Then
\[
| H((\zz - \zz') \opc (\yy - \yy') | \leq \frac8{\log 2} \mathcal{B}(\zz - \zz')
        \sqrt{2d(\yy,\yy')}.
\]
\end{lemma}

We are now ready to tackle the higher order contributions. Let $M_1, M_2, \ldots$ denote constants
independent of the channel.

\begin{proposition}\label{prop-21}
With the notation from the beginning of this section, for any $f \in \mathcal{F}$ (extended by
    linearity), we have
\begin{align*}
\lim_{\dhc\to 0}  \frac{H( \dxx \opc f(\dxx))}{\dhc} &= 0,\\
\lim_{\dhc\to 0}  \frac{H( \dcc \opc f(\dxx))}{\dhc} &= 0.\\
\end{align*}
\end{proposition}
\begin{IEEEproof}
We concentrate on the first limit, as the second is
similar but easier. Applying Lemma \ref{lem-20} we obtain the upper
bound
\[
\lim_{\dhc\to 0} M_1 \frac{\mathcal{B}(\dxx) \sqrt{2d(f(\xx), f(\xx + \dxx))}}{H(\dcc)}.
\]

Since the parametrization is just a linear interpolation between $\cc^*$ and $\Delta_0$ and
$H(\cdot)$ and $\mathcal{B}(\cdot)$ are linear functionals, we have that $H(\dcc) = M_2
\mathcal{B}(\dcc)$.
Then we can replace the
denominator by the Bhattacharyya quantity and use the regularity
condition \eqref{eq-16}. The only thing left to be shown is that
$\sqrt{2d(f(\xx), f(\xx + \dxx))} \to 0$. This follows from inequality \eqref{eq-51}
and the fact that $d$ is a metric.
\end{IEEEproof}

The main tool to turn $\ops$ into $\opc$ and vice-versa is the
following.

\begin{lemma}[Duality lemma, \cite{ruediger-tom-book}] \label{lem-22}
Let $\zz, \zz', \yy, \yy' \in \xspace$. Then
\[
H(\zz \opc \yy) + H(\zz \ops \yy) = H(\zz) + H(\yy).
\]
For differences of densities, because of linearity of $H$, this takes the forms
\begin{align}
H((\zz - \zz')\opc \yy) + H((\zz-\zz') \ops \yy) = H(\zz - \zz'), \label{eq-52} \\
H((\zz - \zz')\opc (\yy-\yy')) + H((\zz-\zz') \ops (\yy-\yy')) = 0. \label{eq-53} 
\end{align}
\end{lemma}

Proposition \ref{prop-21} with the identity map as $f$ and \eqref{eq-53} implies
\begin{align}
\lim_{\dhc\to 0}  \frac{|H( \dxx \ops \dxx)|}{\dhc} = 0.
\end{align}

\begin{proposition}[Proposition 6 in \cite{pfister-nicolas-kumar-young}]\label{prop-23}
If $\zz$ is any symmetric measure (not necessarily signed), then
\[
H(\zz) = \zz(\overline{\mathbb{R}}) - \sum_{k = 1}^\infty \frac{(\log 2)^{-1}}{2k(2k-1)} M_k(\zz),
\]
where $M_k(\zz) = \int (\tanh h)^{2k} d\zz(h)$ and $\zz(\overline{\mathbb{R}})$ is the total mass of
$\zz$.

Moreover, for any symmetric measures $\zz_1$ and $\zz_2$,
    \[
    M_k(\zz_1 \ops \zz_2) = M_k(\zz_1) M_k(\zz_2).
    \]
\end{proposition}

Since the quantities $M_k(\dxx \ops \dxx) = M_k(\dxx)^2$ are all positive, the previous proposition
implies that
\begin{align}\label{eq-55}
|H(\dxx\ops\dxx\ops \yy)| \leq |H(\dxx \ops\dxx)|,
\end{align}
for all $\yy \in \xspace$. By an application of \eqref{eq-52}, one also obtains
\begin{align}\label{eq-56}
|H((\dxx\ops\dxx\ops\yy_1)\opc\yy_2)| \leq 2 |H(\dxx \opc\dxx)|.
\end{align}

We are finally ready to state the result proving the vanishing contribution of higher order terms:
\begin{lemma} \label{lem-24}
We have
\begin{align}\label{eq-57}
&\lim_{\dhc\to 0} \!\!\! \frac{H((\dxx\ops\dxx\ops g_1(\xx,\dxx)) \opc g_2(\xx,\dxx,\cc,\dcc))}{\dhc} \!=\! 0,\\\label{eq-58}
&\lim_{\dhc\to 0} \!\!\! \frac{H((\dxx\ops g_1(\xx)) \opc (\dxx \ops g_2(\xx)) \opc g_3(\xx,\dxx,\cc))}{\dhc} \!=\! 0,\\\label{eq-59}
&\lim_{\dhc\to 0} \!\!\! \frac{H(\dcc \opc (\dxx \ops g_2(\xx)) \opc g_3(\xx,\dxx,\cc))}{\dhc} \!=\! 0.
\end{align}
\end{lemma}

\begin{IEEEproof}
The limit \eqref{eq-57} is a direct consequence of \eqref{eq-56}. The third one, \eqref{eq-59}, is a
consequence of Proposition \ref{prop-21}. The second one can also be reduced to the form appearing
in Proposition \ref{prop-21} by using the Duality Lemma twice:
\begin{align*}
&\quad H((\dxx\ops g_1(\xx)) \opc (\dxx \ops g_2(\xx)) \opc g_3(\xx,\dxx,\cc)) \\
&=      H(\dxx\ops g_1(\xx) \ops ((\dxx \ops g_2(\xx)) \opc g_3(\xx,\dxx,\cc))) \\
&=      H(\dxx\opc (g_1(\xx) \ops ((\dxx \ops g_2(\xx)) \opc g_3(\xx,\dxx,\cc)))).
\end{align*}
\end{IEEEproof}

\section{Conclusions}
%TODO: \addtolength{\textheight}{-12cm}
The present analysis can be extended with almost no change to arbitrary check-node degree distributions whose generating polynomial $P(x) = \sum_{K \leq 0} \rho_K x^K$ is convex for $x \in [-1, 1]$. Experimental evidence suggests that even this condition can be relaxed, but new ideas seem to be required to extend the proofs. A possible route would be to show self-averaging properties for overlap functions, which would allow to use the convexity of $x \mapsto P(x)$ for $x \geq 0$, which holds for any degree distributions (see  \cite{macris-kudekar} for a related approach).

The idea of using spatial coupling as a proof technique potentially goes beyond coding theory. We can use it to analyze the free energy of general spin glass models and find exact characterizations or bounds on their phase transition thresholds. We plan to come back to this problem in a forthcoming publication.

Finally, let us also mention that recently, algorithmic lower bounds to thresholds of constraint-satisfaction problems were derived 
by comparing simple and spatially-coupled constraint-satisfaction models (see  \cite{hamed-nicolas-ruediger}, \cite{dimitris-hamed-nicolas-ruediger}). 

\section*{Acknowledgements}
Andrei Giurgiu acknowledges support from the Swiss National Science Foundation grant No.~200020-140388.  

\appendices
\section{Proof of \eqref{eq-approx-b}} \label{ap-proof-b}
\begin{proposition}
	% TODO: channel realization: check wording
Given a fixed configuration graph $G$ whose underlying type set is
$m$-admissible for $m > K^2$ and a fixed channel realisation $h$, then
with the notation from the proof of Lemma \ref{lem-core} we have that
	\begin{align}
		&\frac{1}{|B_\alpha|} \sum_{a \in B_{\alpha}} \sigrep = \nonumber\\
		&\quad\quad = \frac{1}{|B'_\alpha|} \sum_{a \in
			B'_{\alpha}} \sigrep + O\left( \frac{1}{m}\right).
	\end{align}
\end{proposition}
\begin{IEEEproof}
	Rewrite the left hand side as	
	\begin{align}\label{eq-ae1}	
		%\frac{1}{|B_\alpha|} \sum_{a \in B_{\alpha}} \sigrep &= 
		\mkern-12mu
		\frac{1}{|B'_\alpha|}\frac{|B'_\alpha|}{|B_\alpha|} \Bigg( \!\sum_{a
			  \in B'_{\alpha}} \!\!\sigrep - \!\!\!
			  \!\!\!\sum_{a \in B'_\alpha \setminus B_\alpha}\!\!\!\!\! \sigrep \!\!\!\Bigg).
	\end{align}
	
	We will first find an estimate of the quantity $|B'_{\alpha}\setminus
	B_\alpha|$, i.e. the number of (pseudo-)check constraints that connect to
	at least one socket multiple times. To do this, let us look at the subset
	of $B'_\alpha$ where $a_i = a_j$ (i.e. edges $i$ and $j$ connect to the same socket), for some
	distinct $i,j$ with $1 \leq i,j \leq K$. The cardinality $q_{i,j}$ of this subset is $0$ if
	$\alpha_i \neq \alpha_j$, and is equal to $|B'_\alpha| /
|F_i| \leq |B'_\alpha| / m$ if $\alpha_i = \alpha_j$. 

	A (rough) upper bound for $|B'_{\alpha} \setminus B_\alpha|$ is given then by sum
	$\sum_{i \neq j} q_{i,j}$, which in turn never exceeds $K^2
	|B'_\alpha| / m$.

	\vspace{0.1cm}

	We are now able to bound the ratio $|B'_\alpha|/|B_\alpha|$ appearing in
	\eqref{eq-ae1} by $m / (m - K^2)$. Indeed, this follows from 
	\[\frac{|B'_\alpha|}{|B_\alpha|} = \frac{|B'_\alpha|}{|B'_\alpha| - |B'_\alpha \setminus B_\alpha|}.\]
		
	The absolute value of the second sum in \eqref{eq-ae1}
	is clearly upper-bounded by $|B'_{\alpha} \setminus B_\alpha|$, since the
	bracket takes values between $0$ and $1$.
	Putting everything together, we obtain 
	\begin{align*}
		&\frac{1}{|B_\alpha|} \sum_{a \in B_{\alpha}} \sigrep 
		\leq\nonumber\\&\quad\quad\leq 
		\left( \frac{m}{m-K^2}\right)\frac{1}{|B'_\alpha|} \sum_{a \in
			B'_{\alpha}} \sigrep + \frac{K^2}{m-K^2}, \\
		&\frac{1}{|B_\alpha|} \sum_{a \in B_{\alpha}} \sigrep 
		\geq\nonumber\\&\quad\quad\geq 
		\frac{1}{|B'_\alpha|} \sum_{a \in B'_{\alpha}} \sigrep \
			- \frac{K^2}{m-K^2}.
	\end{align*}
\end{IEEEproof}

\bibliographystyle{ieeetr}
\bibliography{interp}

\end{document}